# Pattern Recognition of Aluminium Arbitrage in Global Trade Data

MUHAMMAD SUKRI BIN RAMLI
Asia School of Business
Kuala Lumpur, Malaysia
Email: m.binramli@sloan.mit.edu

**Abstract**

As the global economy transitions toward decarbonization, the aluminium sector has become a focal point for strategic resource management. While policies such as the Carbon Border Adjustment Mechanism (CBAM) aim to reduce emissions, they have inadvertently widened the price arbitrage between primary metal, scrap, and semi-finished goods, creating new incentives for market optimization. This study presents a unified, unsupervised machine learning framework to detect and classify emerging trade anomalies within UN Comtrade data (2020–2024). Moving beyond traditional rule-based monitoring, we apply a four-layer analytical pipeline utilizing Forensic Statistics, Isolation Forests, Network Science, and Deep Autoencoders. Contrary to the hypothesis that Sustainability Arbitrage would be the primary driver, empirical results reveal a contradictory and more severe phenomenon of Hardware Masking. Illicit actors exploit bi-directional tariff incentives by misclassifying scrap as high-count heterogeneous goods to justify extreme unit-price outliers of >$160/kg, a 1,900% markup indicative of Trade-Based Money Laundering (TBML) rather than commercial arbitrage. Topologically, risk is not concentrated in major exporters but in high-centrality Shadow Hubs that function as pivotal nodes for illicit rerouting. These actors execute a strategy of Void-Shoring, systematically suppressing destination data to Unspecified Code to fracture mirror statistics and sever forensic trails. Validated by SHAP (Shapley Additive Explanations), the results confirm that price deviation is the dominant predictor of anomalies, necessitating a paradigm shift in customs enforcement from physical volume checks to dynamic, algorithmic valuation auditing.

## 1. Introduction

As the global economy accelerates its transition toward decarbonization, the aluminium sector has emerged as a focal point for strategic resource management and industrial policy. However, policies such as CBAM have inadvertently widened the price arbitrage between primary metal, scrap, and semi-finished goods. While intended to curb carbon leakage, recent scholarship suggests that such unilateral environmental policies can displace emissions rather than reduce them, creating new incentives for market optimization and trade rerouting (Eicke et al., 2021; Magacho et al., 2024). This study contends that these regulatory pressures are not merely altering trade volumes but are actively reshaping the topology of global trade fraud, necessitating a shift from traditional monitoring to advanced pattern recognition.

**Figure 1. The System Dynamics of Trade Data Evasion vs. Enforcement.**

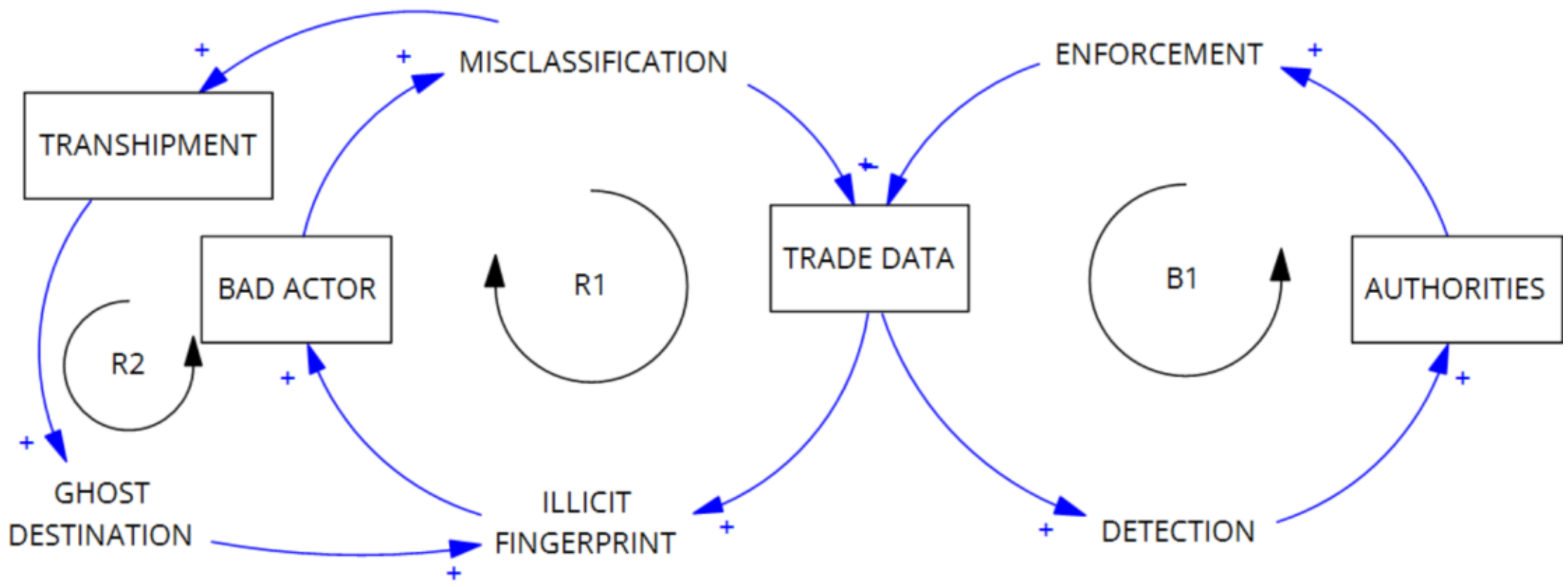

Source: Processed by Author (2025)

To guide the understanding, we draw the causal loops that uncover that the system dynamics of TBML represent a continuous structural tension between criminal concealment and regulatory detection. The system is primarily driven by a reinforcing profit loop (R1) where "Bad Actors" utilize "Misclassification" specifically "Hardware Masking" to over-invoice goods, thereby injecting fraudulent value into "Trade Data" and creating a statistical "Illicit Fingerprint" of massive price markups. To obscure this activity, these actors simultaneously operate a secondary evasion loop (R2) that utilizes "Transhipment" hubs to mask final delivery, creating "Ghost Destinations" where exports leave the origin but effectively vanish from import records. Counteracting these reinforcing forces is the balancing regulatory loop (B1), where "Authorities" leverage advanced "Detection" to identify these price outliers and broken data trails; successful detection triggers "Enforcement" measures that seize assets and restrict manipulated flows, applying the necessary pressure to dampen the fraud and stabilize the trade system.

The primary driver of the observed trade anomalies is the 'tax wedge' the profitable margin created by the discrepancy in financial liability between different classifications of aluminium products. As established in the literature on value-added tax (VAT) evasion, distinct differentials in tax rates between jurisdictions or product categories create structural arbitrage opportunities (Keen & Smith, 2006). This creates a distinct financial hierarchy within the global tariff structure: while raw materials and scrap often enjoy duty-free status upon import, finished goods typically face significantly higher duties. This structure mirrors the 'Missing Trader' fraud mechanics described by Ainsworth (2010), where illicit actors exploit these administrative gaps to capture the margin between duty-free inputs and taxed outputs.

This arbitrage opportunity is visually quantified in the provided financial analyses. Figure 1 highlights the import landscape, where "Top Risk" finished goods face substantial tax burdens compared to the 0% duty on scrap and raw materials, incentivizing traders to fraudulently declare finished goods as raw materials to evade import taxes. Conversely, Figure 2 reveals a counter-incentive in the export market, where "Waste & Scrap" faces a distinct 10% export duty while other processed articles are duty-free (NIL). This complex environment creates a bi-directional motive for misclassification: traders are incentivized to mislabel imports as scrap to avoid tariffs, yet must mislabel scrap exports as finished goods to avoid export duties. As noted by the OECD (2019), such tariff differentials frequently exacerbate distortions in the aluminium value chain, confirming that specific tariff lines are not merely administrative categories but "Top Risk" vectors for fiscal evasion.

**Figure 2. Comparative Analysis of Import Duties and Tax Burdens by HS Code in Malaysia.**

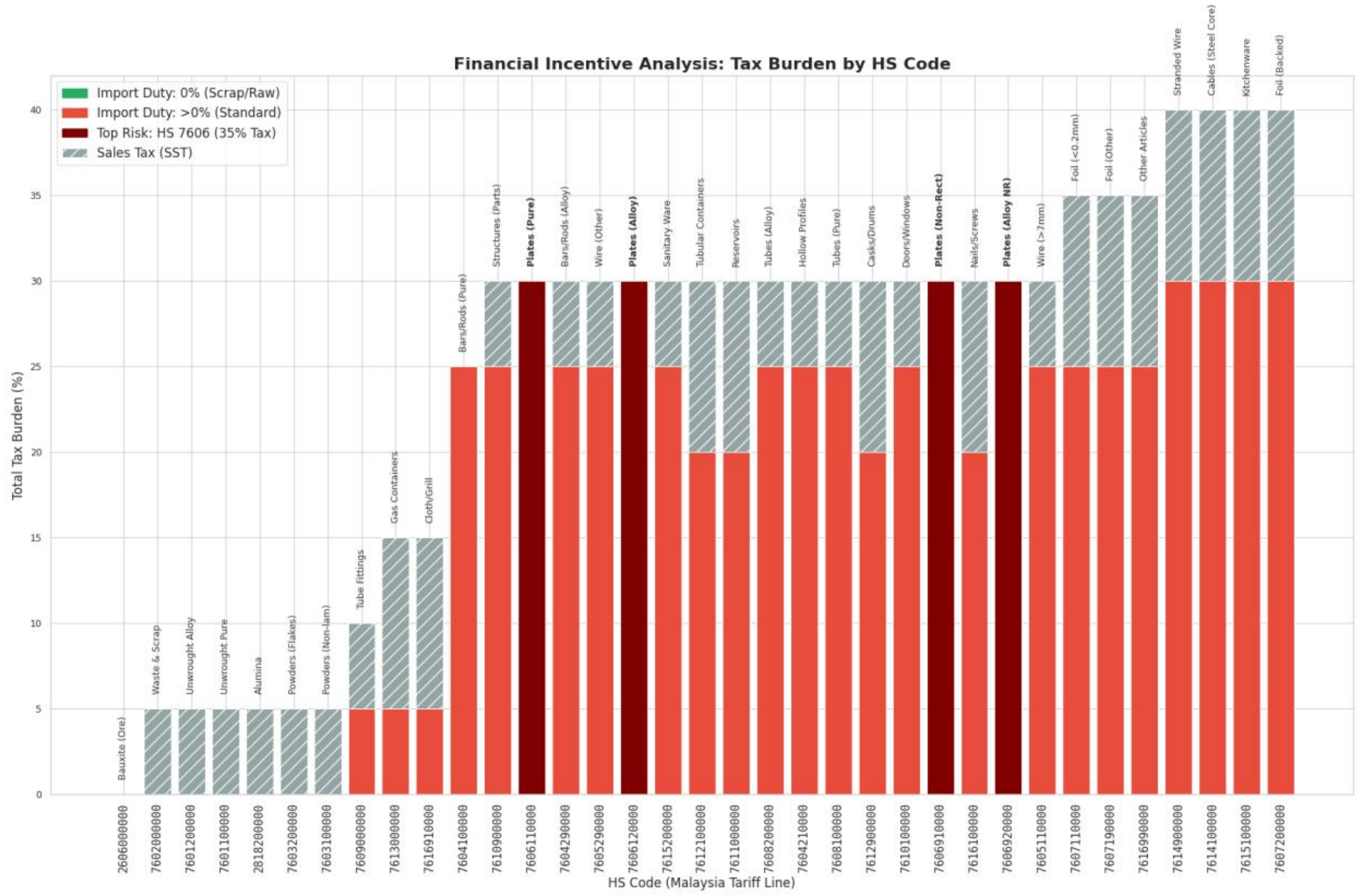

Source: Processed by Author (2025)

The correlation between these incentive structures and actual illicit activity is not static. Figure 4 presents a time-series analysis comparing total global trade volume against detected anomalies from 2020 to 2024, revealing that trade anomalies are highly responsive to broader market shifts. The data indicates that as global trade values fluctuate, seen in the sharp decline and subsequent stabilization in the chart the frequency of anomalies adapts in tandem. This suggests that these irregularities are statistically significant signals rather than random noise. This behavior aligns with Buehn and Eichler’s (2011) findings on trade mis-invoicing, which suggest that illicit actors actively optimize their strategies to "dodge" macroeconomic shifts and regulatory barriers, effectively utilizing the global trade system as a mechanism for arbitrage.

**Figure 3. Comparative Analysis of Export Duties and Tax Burdens by HS Code in Malaysia.**

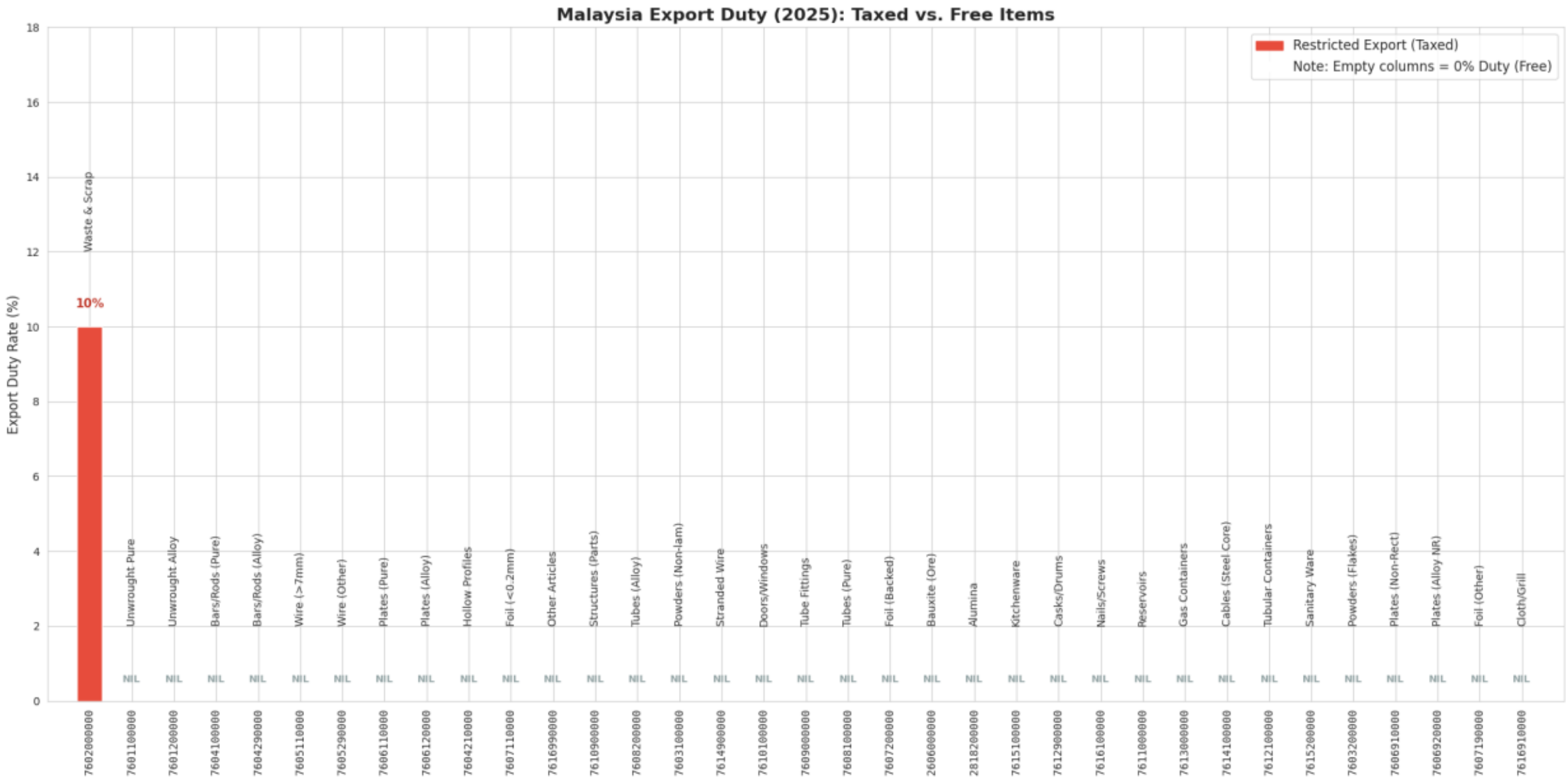

Source: Processed by Author (2025)

**Figure 4. Time Series Analysis of Global Trade Volume vs. Detected Anomalies.**

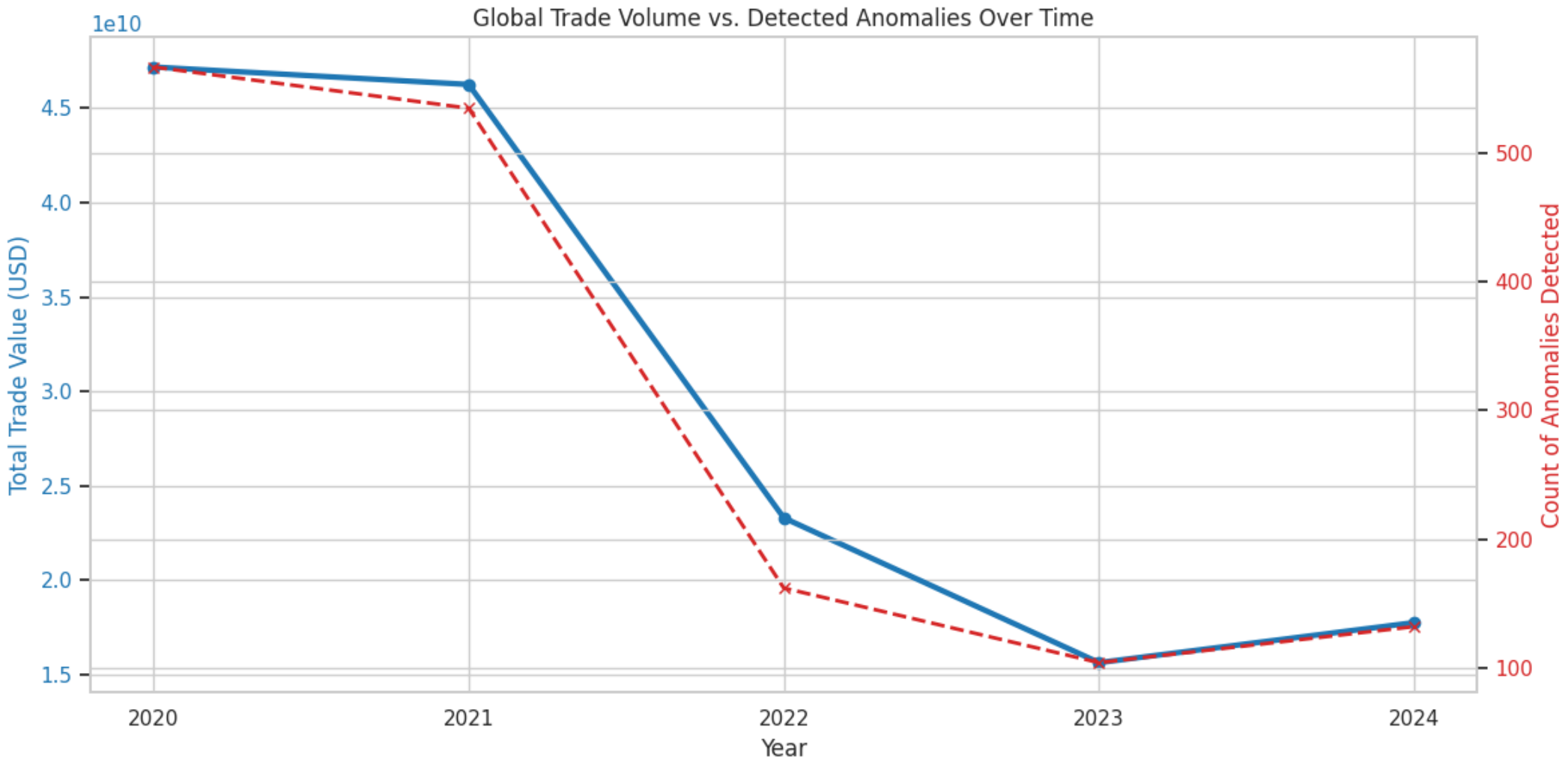

Source: Processed by Author (2025)

**Figure 5. Time Series Analysis of Global Trade Volume vs. Detected Anomalies.**

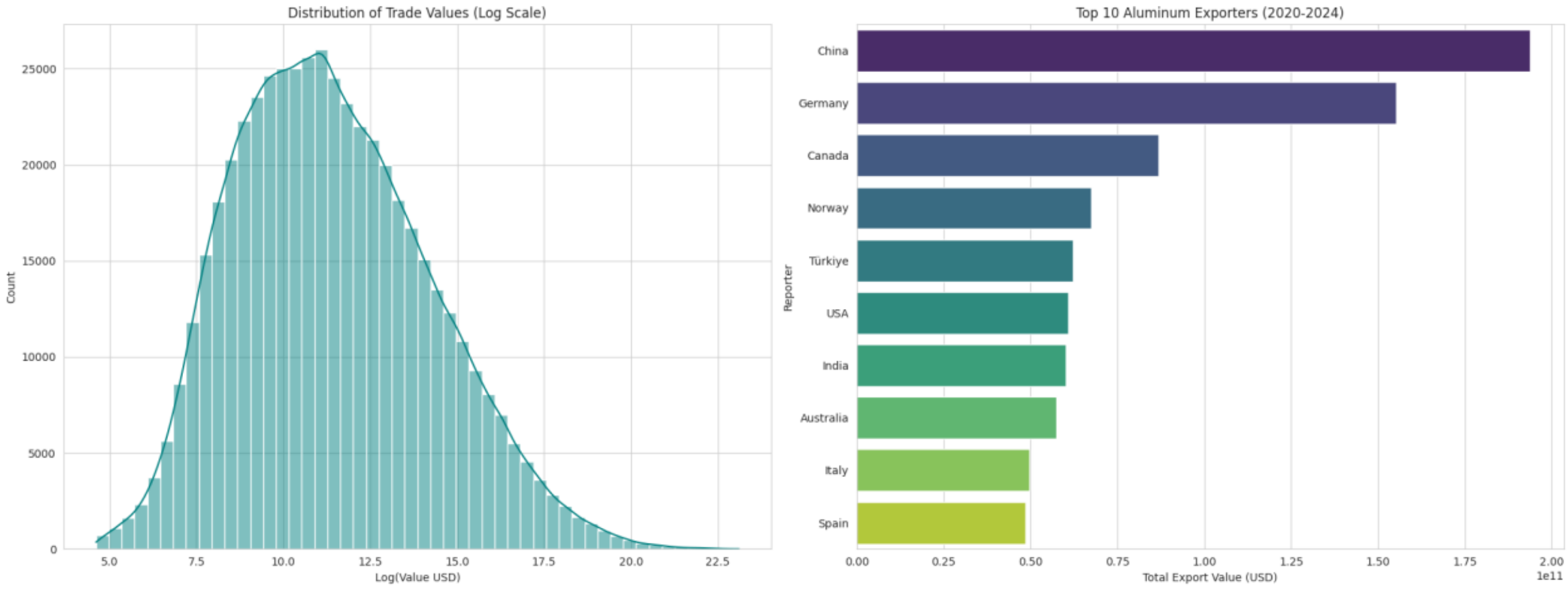

Source: Processed by Author (2025)

To accurately detect these deviations, it is necessary to first establish the statistical baseline of "normal" trade. Figure 5 provides this "ground truth" through a composite analysis of the global market. The left-hand histogram displays the log-scale distribution of trade values, forming a bell curve that mathematically defines the standard value range for legitimate transactions. Simultaneously, the right-hand chart ranks the top 10 exporters by total export value, identifying China and Germany as the dominant market leaders. By visualizing the standard distribution and key players of the global aluminium market, this figure serves as the control group for the experiment, allowing for the subsequent unsupervised learning algorithms to distinguish between standard market behavior and the outlier's indicative of fraud.

**Figure 6. Hierarchical Taxonomy of Aluminium HS Codes within the Research Scope.**

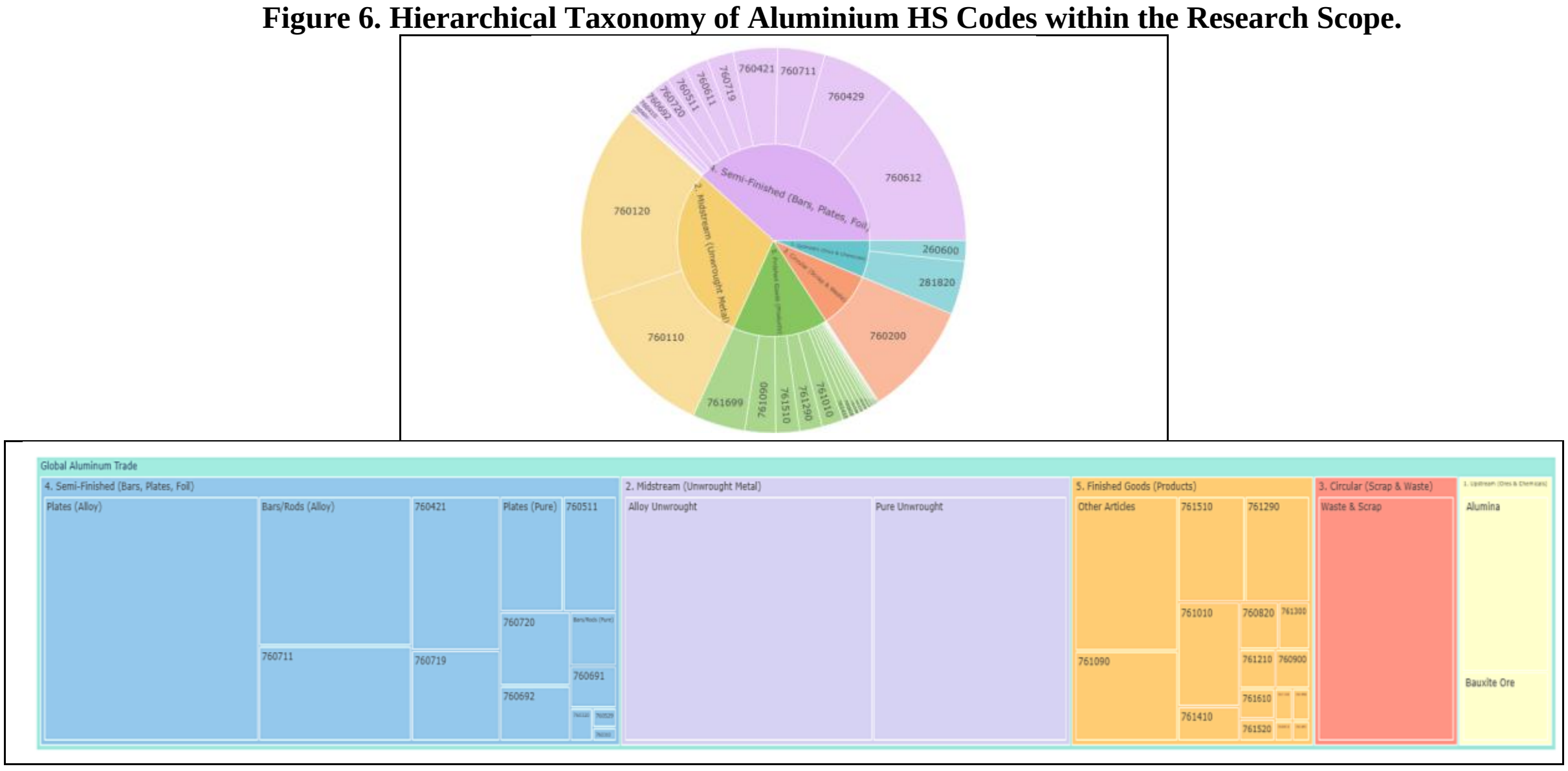

Source: Processed by Author (2025)

Figure 6 serves as the strategic map for this investigation, organizing the complex and often chaotic list of Harmonized System (HS) codes into a structured, hierarchical taxonomy. This sunburst visualization strictly defines the boundaries of the products analyzed in the research by segmenting them into clear logical categories: "Upstream" (Ores & Chemicals), "Midstream" (Unwrought Metal), "Downstream" (Semi-Finished goods like plates and foil), and "Circular" (Scrap & Waste). This visualization is essential for the study's scope because it visualizes the specific product boundaries that fraudsters often attempt to blur. By visually separating raw materials, which typically enjoy low or zero tariffs, from finished goods that face high tariffs, the chart highlights the "risk zones" where misclassification is most likely to occur to capture tax margins.

**Figure 7. Treemap of Global Trade Structure by Volume and Unit Price.**

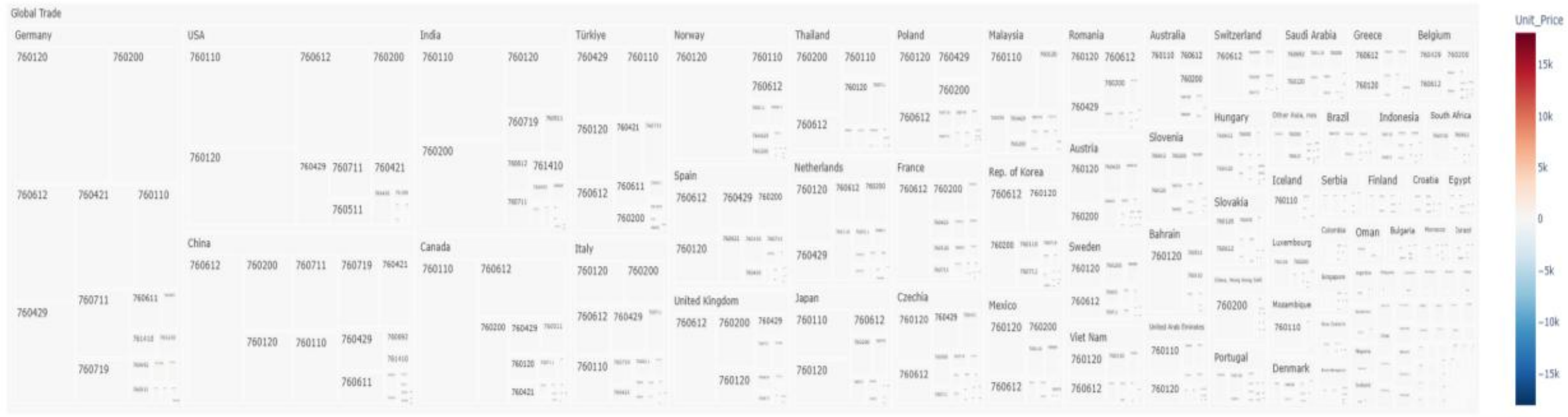

Source: Processed by Author (2025)

Complementing this categorical breakdown, Figure 7 provides a dense, macro-level overview of the global aluminium economy through a treemap visualization where the size of each tile represents trade volume and the color intensity represents unit price. The primary function of this chart is to reveal the "economic weight" of different market segments and major trading regions at a glance. By mapping these dimensions simultaneously, the researcher can instantly differentiate between high-volume, low-value commodities, such as raw scrap shown in blue and low-volume, high-value niches, such as specialized alloys shown in red. This establishes a structural baseline for the market, allowing the study to distinguish where value is legitimately concentrated versus where it may be artificially inflated, serving as a prerequisite for identifying pricing anomalies.

## 2. Theoretical Framework

### 2.0 Typologies of Trade Arbitrage

To systematically detect anomalies, this study categorizes trade fraud into three distinct theoretical typologies: Sustainability Arbitrage, Fiscal Asymmetry, and Geopolitical Rerouting. Each typology represents a specific economic behavior driven by the "tax wedge" and regulatory pressures previously outlined.

### 2.1 Sustainability Arbitrage (The "Greenwashing" Incentive)

The transition to a circular economy has fundamentally altered the pricing dynamics of raw materials. As noted by Söderholm and Ekvall (2020), policy drivers are increasingly incentivizing metal recycling, creating a premium for sustainable inputs. This gives rise to 'Sustainability Arbitrage,' where the market incentivizes the fraudulent mislabelling of high-carbon primary metal as exempt 'green' recycled materials to evade carbon tariffs. This behavior is driven by the 'Greenium' a tangible price premium for sustainable, low-carbon materials. As noted by McKinsey & Company (2022), the demand for these low-carbon inputs is rapidly outpacing supply, creating a structural price divergence that illicit actors can exploit. This aligns with Aegis Hedging’s (2020) analysis of green aluminium pricing, which predicts that this spread will become a primary driver of commodity cost differentials in the coming decade.

**Figure 8. Forecasted Price Convergence of Prime and Scrap Aluminium (2020–2026).**

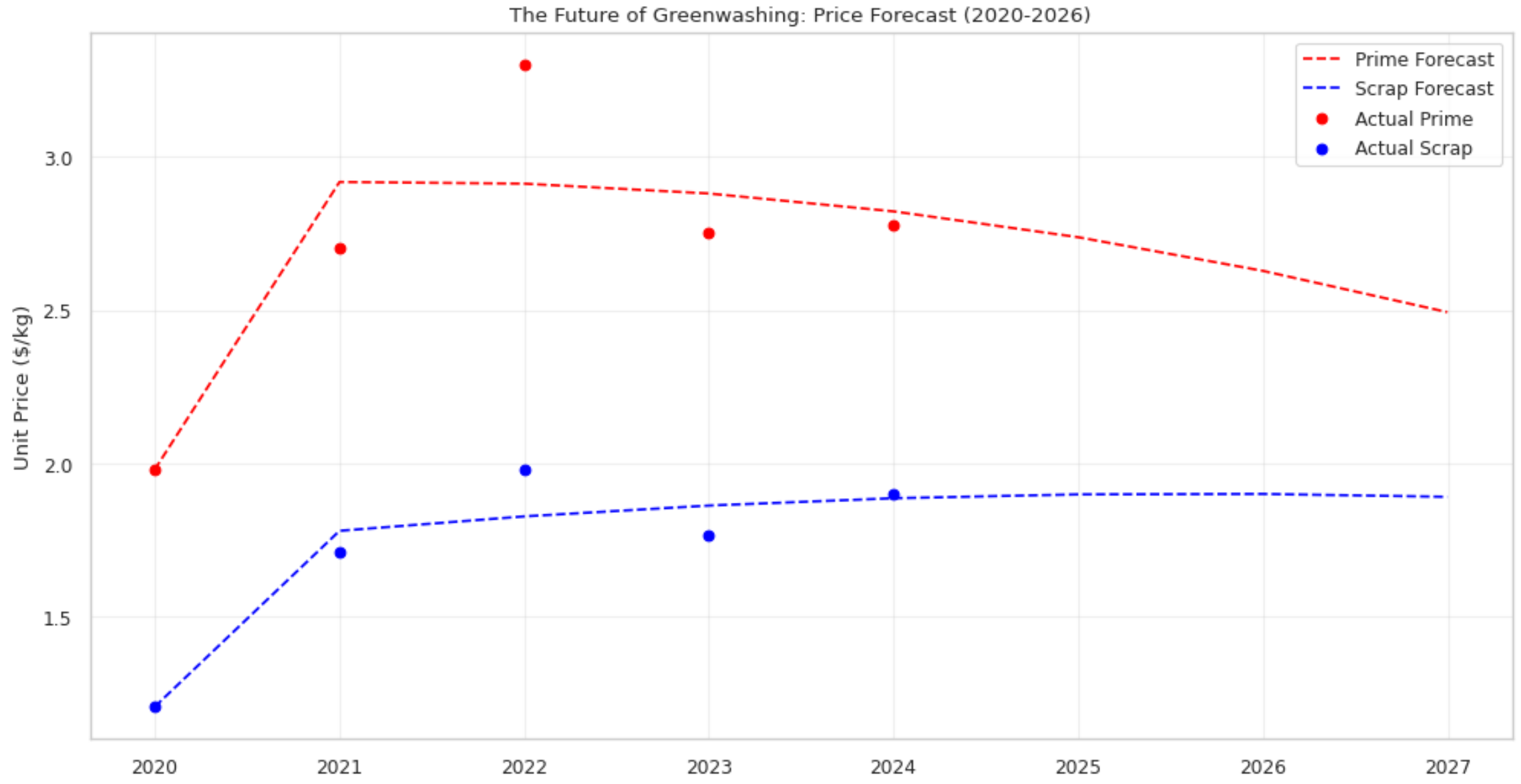

Source: Processed by Author (2025)

Figure 8 illustrates this shifting dynamic through a forecasted price convergence of Prime and Scrap aluminium (2020–2026). The chart predicts that the historically lower price of scrap (indicated by the blue dotted line) will rise to meet primary metal prices (the red dotted line). It is critical to delineate the metallurgical distinction between primary (mined) aluminium and secondary (recycled) flows, as these are not perfectly fungible assets. Unlike primary aluminium, which is refined directly from bauxite ore to achieve near-perfect purity, recycled aluminium inevitably accumulates alloying elements and impurities during its lifecycle that are technically difficult to remove.

However, despite these chemical differences, the two forms are visually indistinguishable to the naked eye. This opacity creates a vulnerability for "Sustainability Arbitrage," particularly under regimes like the CBAM. Illicit actors can exploit this by remelting carbon-intensive primary ingots into non-standard shapes or semi-finished forms to disguise their origin, effectively "laundering" the metal's carbon footprint. Because the physical product looks identical, these actors can fraudulently mislabel high-carbon prime metal as exempt recycled material, capturing the "green" premium while bypassing environmental tariffs.

This theoretical forecast is validated by the empirical data in Figure 9, which shows that the median unit prices for HS 7601 (Prime) and HS 7602 (Scrap) have physically converged over the last four years. Rather than signaling a decline in market utility, this convergence indicates that low-carbon scrap has evolved into a critical strategic asset. State and non-state actors are now actively competing to stockpile and 'keepsake' recycled aluminium within domestic borders, utilizing it as a defensive shield to lower industrial emissions and mitigate CBAM liabilities. However, as this resource nationalism restricts the global supply of genuine scrap, the Greenium, the pricing of climate risk described by Alessi et al. (2019) intensifies.

**Figure 9. Convergence of Median Unit Prices for Prime vs. Scrap Aluminium (2020–2024).**

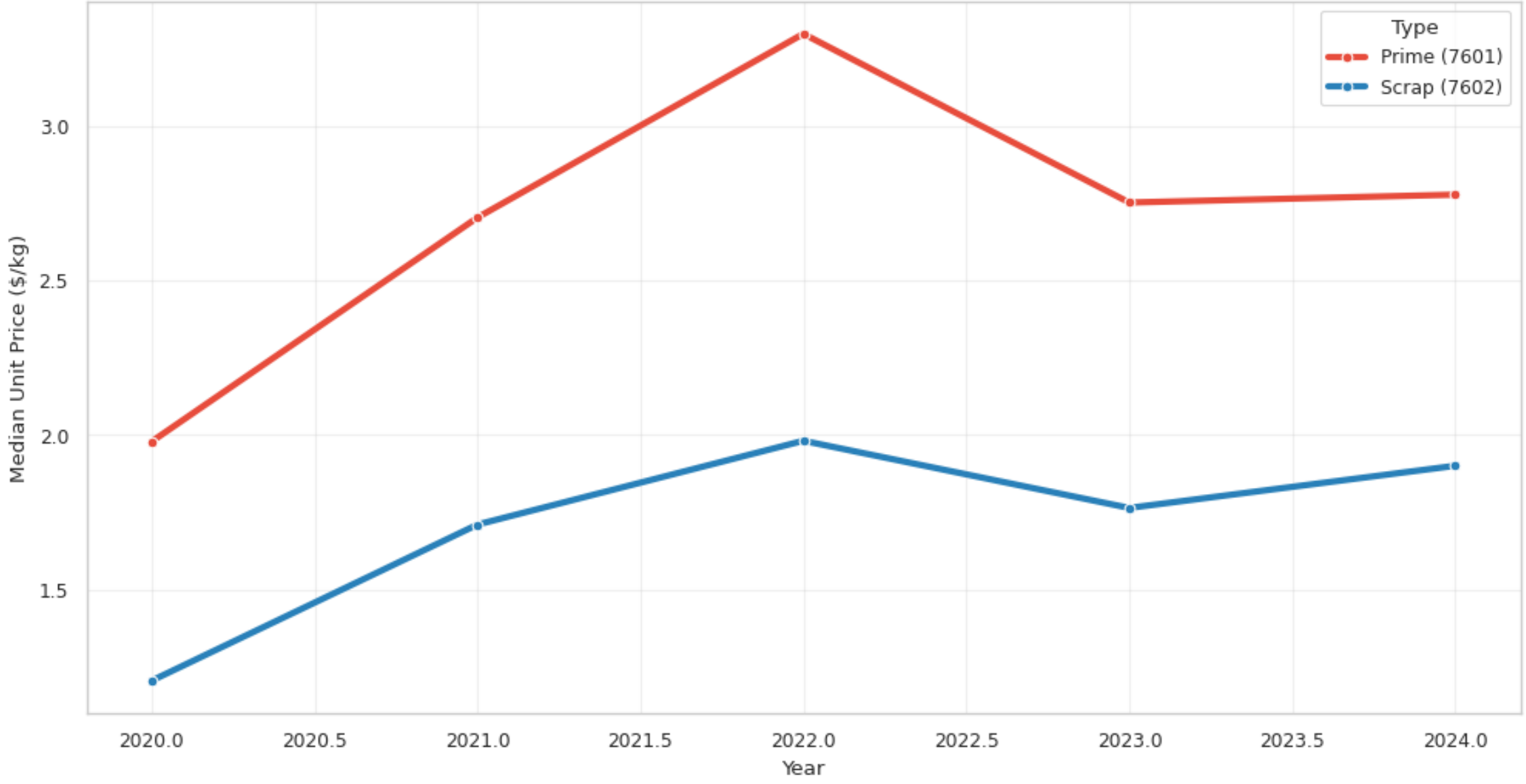

Source: Processed by Author (2025)

### 2.2 Fiscal Asymmetries ("Phantom Trade")

The second typology addresses systemic data discrepancies, often referred to in the literature as "missing imports". Fisman and Wei (2004) established that tax rates correlate directly with evasion, evidenced by gaps where reported exports exceed reported imports. While Buehn and Eichler (2011) categorize this as 'trade mis-invoicing', a deliberate strategy to dodge capital controls, our analysis reveals this to be a fundamental structural failure which we term a "Mirror Fracture". This phenomenon is visible on a macro scale in Figure 10, where global exports (blue line) consistently outpace imports (red line), leaving a massive "Missing Metal Gap" represented by the grey bars below the baseline. Unlike random noise or simple administrative error, a Mirror Fracture represents a coordinated effort to shatter the "Perfect Symmetry" required for transparent monitoring. This deliberate statistical breach creates a jagged data landscape where high-volume flows exit the export ledger but fail to materialize in import records, effectively serving as the primary camouflage for fiscal asymmetry and smuggling operations.

**Figure 10. Convergence of Median Unit Prices for Prime vs. Scrap Aluminium (2020–2024).**

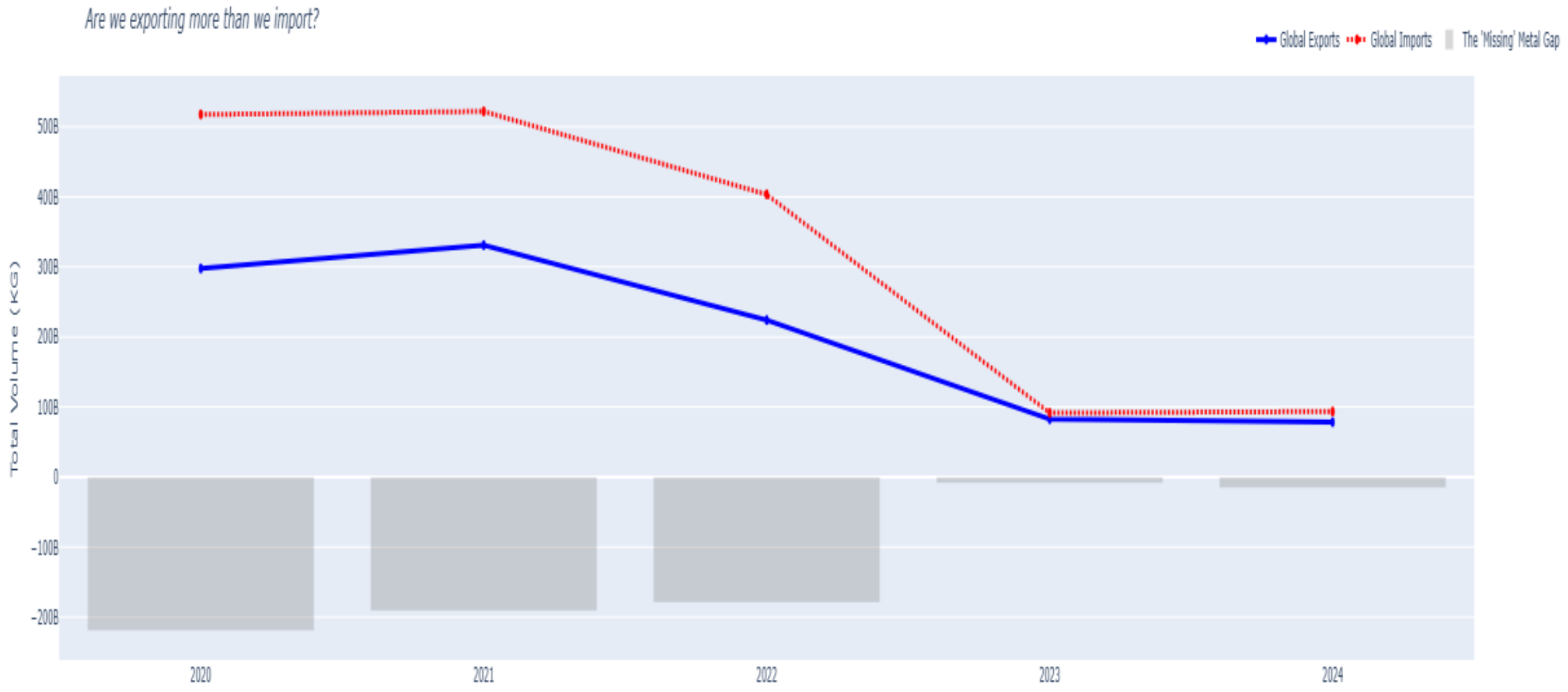

Source: Processed by Author (2025)

Furthermore, Figure 11 quantifies the severity of this trend, showing a steep downward trajectory in the "Gap Percentage" (the red line). This indicates that data transparency is actively deteriorating over time. This widening "Black Hole" aligns with the theoretical analysis of smuggling by Bhagwati and Hansen (1973), which posits that as illicit trade becomes institutionalized, official statistics increasingly decouple from physical reality.

**Figure 11. Temporal Analysis of Missing Trade Volumes and Reporting Gap Percentages.**

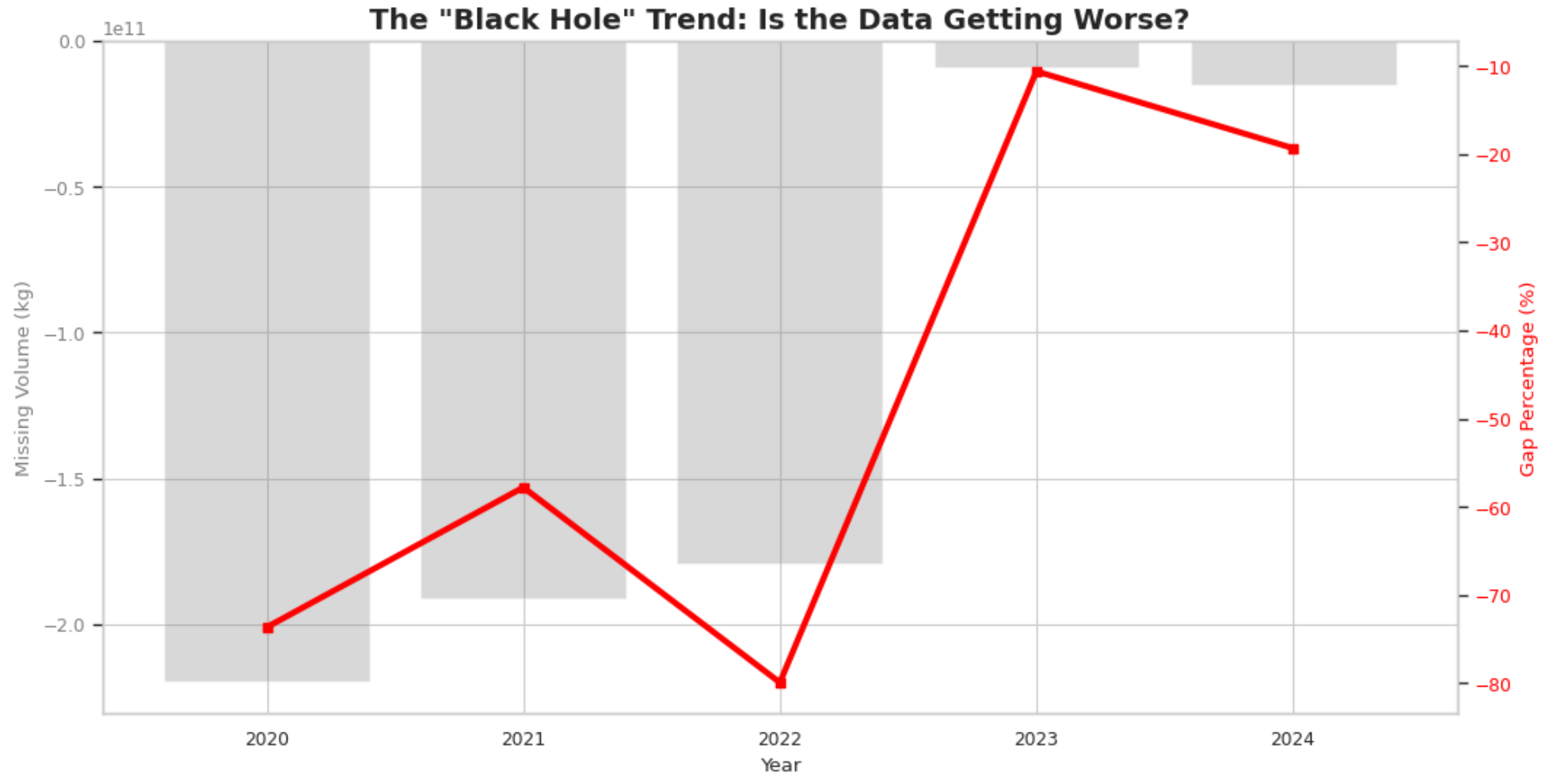

Source: Processed by Author (2025)

### 2.3 Geopolitical Rerouting (Transshipment and Sanctions Evasion)

The third typology concerns the physical displacement of trade in response to regulatory barriers. Reuter (2014) describes this as the "balloon effect" in illicit markets, where enforcement measures in one jurisdiction simply displace the trade activity to another. This phenomenon is particularly relevant for sanctions evasion, as analyzed by Wesselbaum and Sanders (2023) in the context of military goods. Figure 12 simulates this effect within the aluminium sector by modeling a hypothetical ban by a major North American economy on East Asian aluminium. The simulation results predict that restricted trade volumes would not vanish; instead, they would shift to intermediaries, with Southern European nations and neighboring North American partners identified as the top predicted destinations. This "Geopolitical Rerouting" creates complex transshipment anomalies, often exacerbated by the "Rotterdam Effect," where major regional ports inflate trade statistics by acting as quasi-transit hubs rather than final destinations (Mellens & Gorter, 2018).

**Figure 12. Projected Trade Displacement Following a Hypothetical North American Economy on East Asian.**

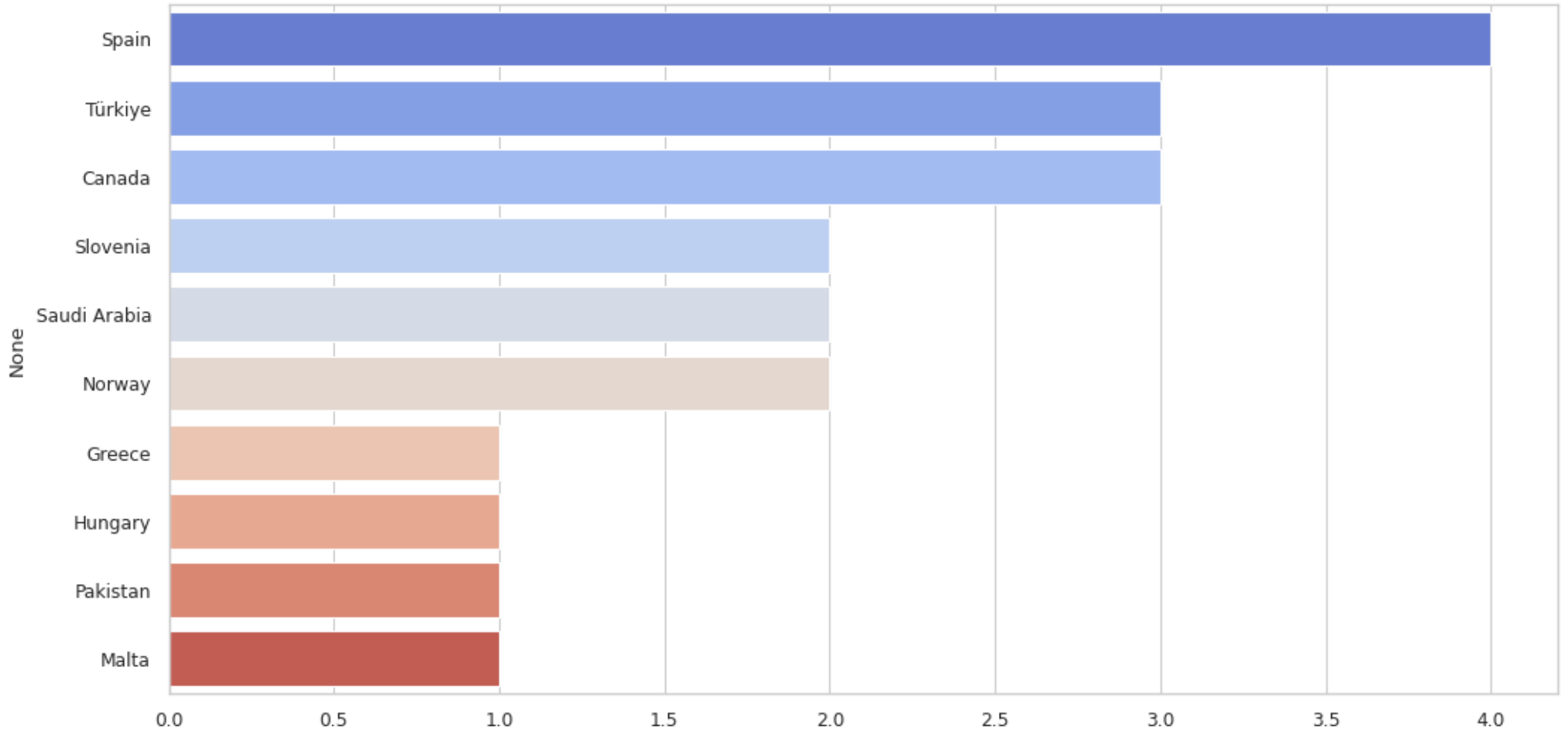

Source: Processed by Author (2025)

## 3. Methodology

### 3.0 Multi-Layered Analytical Pipeline

A Multi-Layered Analytical Pipeline To address the complexity of modern trade fraud, this study moves beyond traditional rule-based monitoring by implementing a four-layer unsupervised machine learning framework applied to official UN Comtrade data. The analytical scope encompasses the entire aluminium value chain, tracking 31 Harmonized System (HS) codes across five production stages to monitor value transformation. These include Upstream raw materials (Bauxite 260600, Alumina 281820), Midstream unwrought aluminium (760110, 760120), and Circular scrap (760200), as well as Semi-Finished inputs (11 codes covering bars, wire, plates/sheets, and foil) and Finished Goods (15 codes ranging from tubes and structures to household articles). Leveraging this granular dataset, the pipeline is designed to validate data integrity, detect volumetric outliers, map structural irregularities, and isolate non-linear value anomalies.

### 3.1 Data Validation: Forensic Statistics (Layer 1)

The first layer serves as a forensic checkpoint to distinguish between systemic data characteristics and specific reporting irregularities. We apply Benford's Law, utilizing the goodness-of-fit testing framework proposed by Barabesi et al. (2018) specifically for customs fraud detection. As detailed by Nigrini (2019) in the context of forensic accounting, legitimate transaction data should naturally conform to a specific logarithmic distribution of leading digits. Consequently, significant deviations from this mathematical curve are not random; they signal potential manipulation where human actors have artificially altered values to evade specific thresholds.

**Figure 13. Benford's Law Validation for Aggregate Global Data Integrity.**

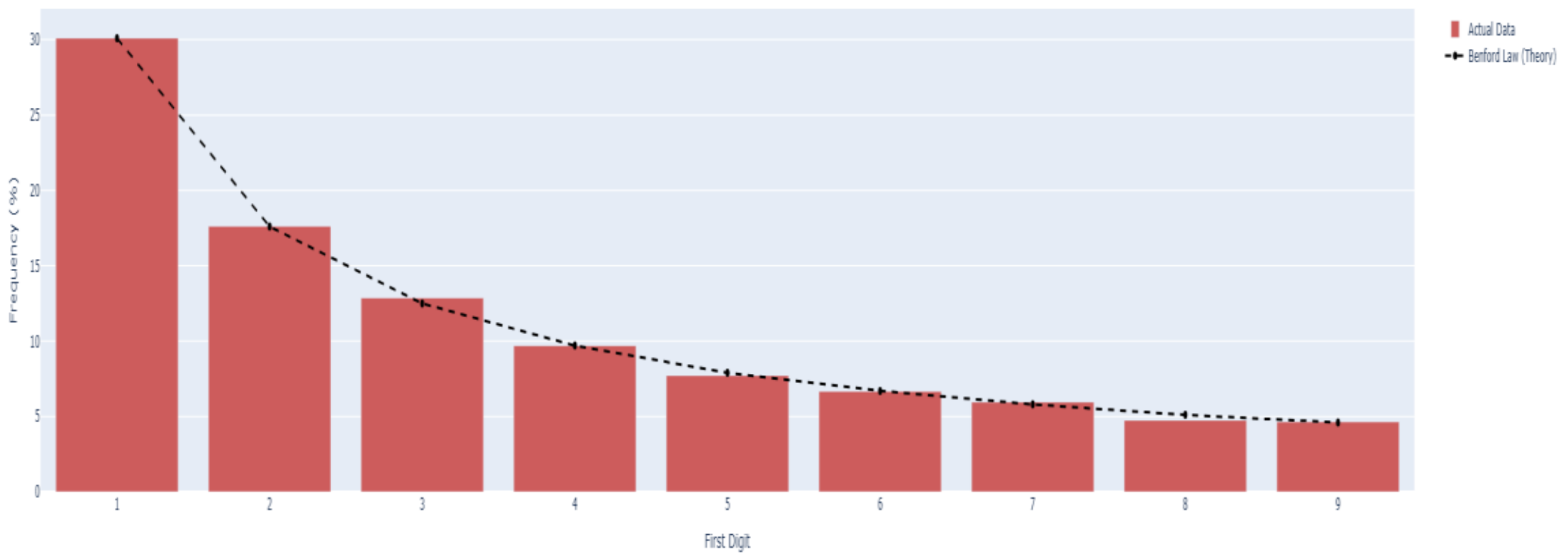

Source: Processed by Author (2025)

**Figure 14. Benford's Law Analysis of East Asian Export Value Distribution.**

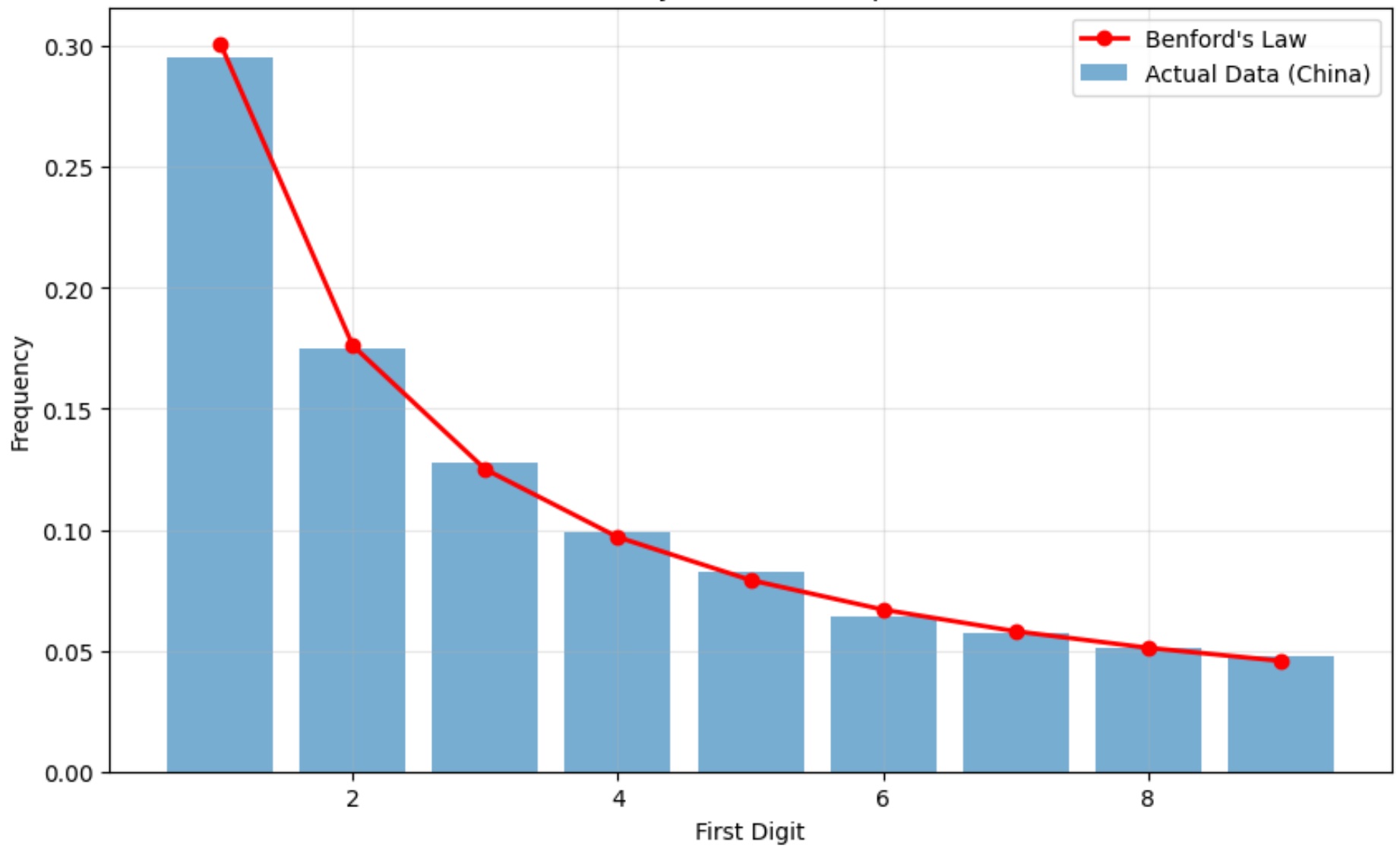

Source: Processed by Author (2025)

As shown in the "Forensic Integrity Check" (Figure 13), the aggregate global dataset largely conforms to the expected logarithmic curve (represented by the black dotted line), validating the macro-level integrity of the data. However, distinct "statistical fractures" emerge when analyzing specific high-risk regions. The "Forensic Analysis: East Asian Export Values" (Figure 14) reveals a significant divergence where the actual data (blue bars) visibly defies the Benford prediction (red line). This lack of fit, manifesting as an unnatural frequency of specific leading digits indicates that the data is not the result of organic commercial activity. These mathematical deviations provide the initial statistical justification for focusing subsequent algorithms on these specific trade routes, effectively flagging them as high-probability vectors for fraud.

### 3.2 Volumetric Anomaly Detection (Layer 2)

The second layer utilizes the Isolation Forest algorithm, originally introduced by Liu et al. (2008). Unlike standard statistical methods that attempt to model what "normal" data looks like, Isolation Forest explicitly isolates anomalies by exploiting the property that they are few and different. It works by randomly selecting a feature and then randomly selecting a split value; anomalies are easier to "isolate" with fewer random partitions than normal data points. This approach is particularly effective for high-dimensional trade datasets.

**Figure 15. Unsupervised Classification of Suspicious Trade Flows vs. Normal Trade.**

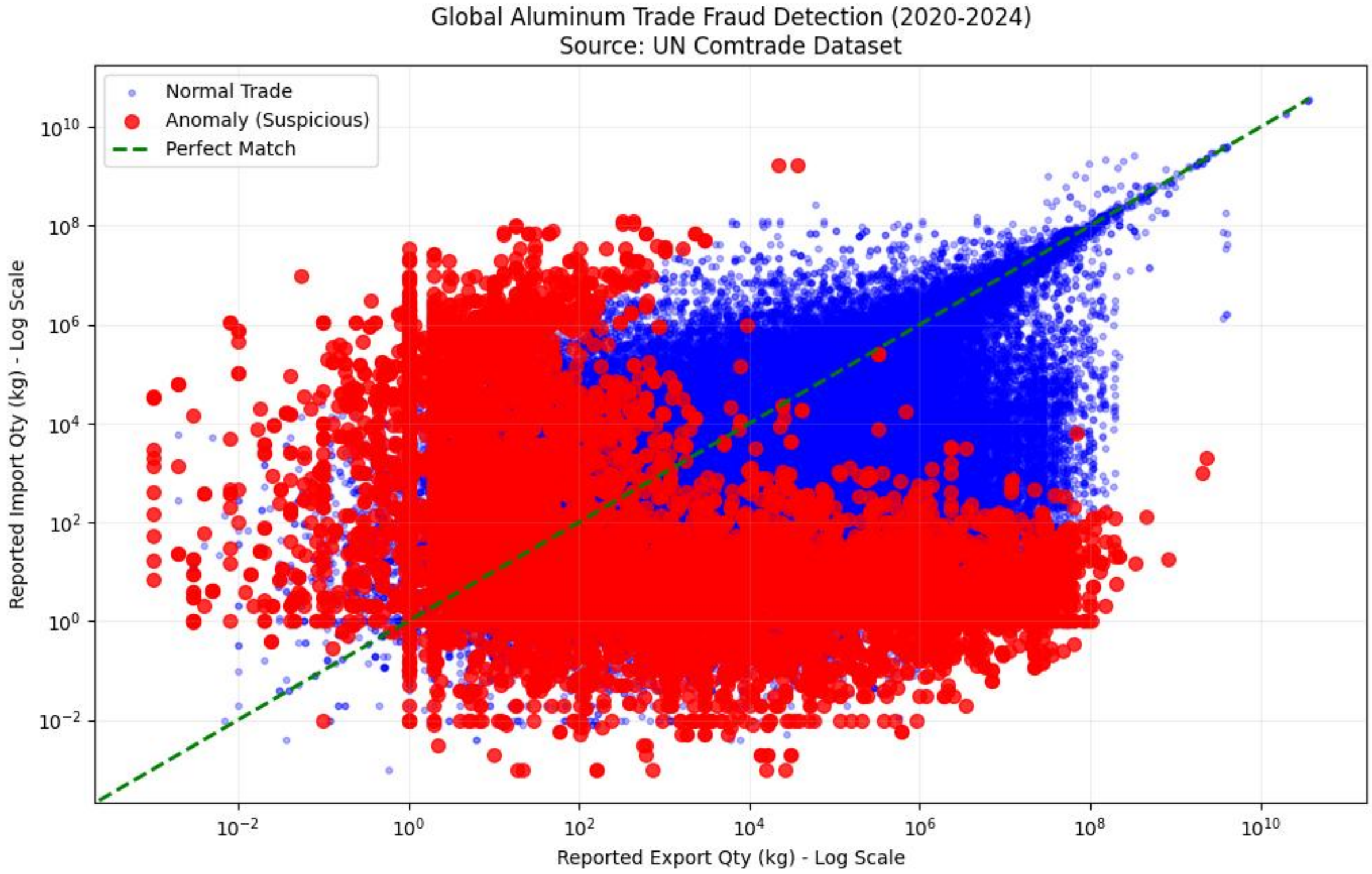

Source: Processed by Author (2025)

The chart in Figure 15 visualizes the output of this classification by plotting reported import against export quantities. The chart maps a clear "decision boundary," where the red dots represent "Suspicious Anomalies" that deviate from the standard market operating patterns (the blue dots) and the perfect match line (green). To address the "black box" nature of machine learning, the "Why is it an Anomaly?" chart (Figure 16) provides a Decision Tree visualization. This explicates the specific logical rules-such as specific "Price Disparity" thresholds or "Gap Ratios" used to flag these transactions. This converts opaque algorithmic detection into actionable, rule-based alerts for customs officials.

**Figure 16. Decision Tree Rules for Interpretable Anomaly Classification.**

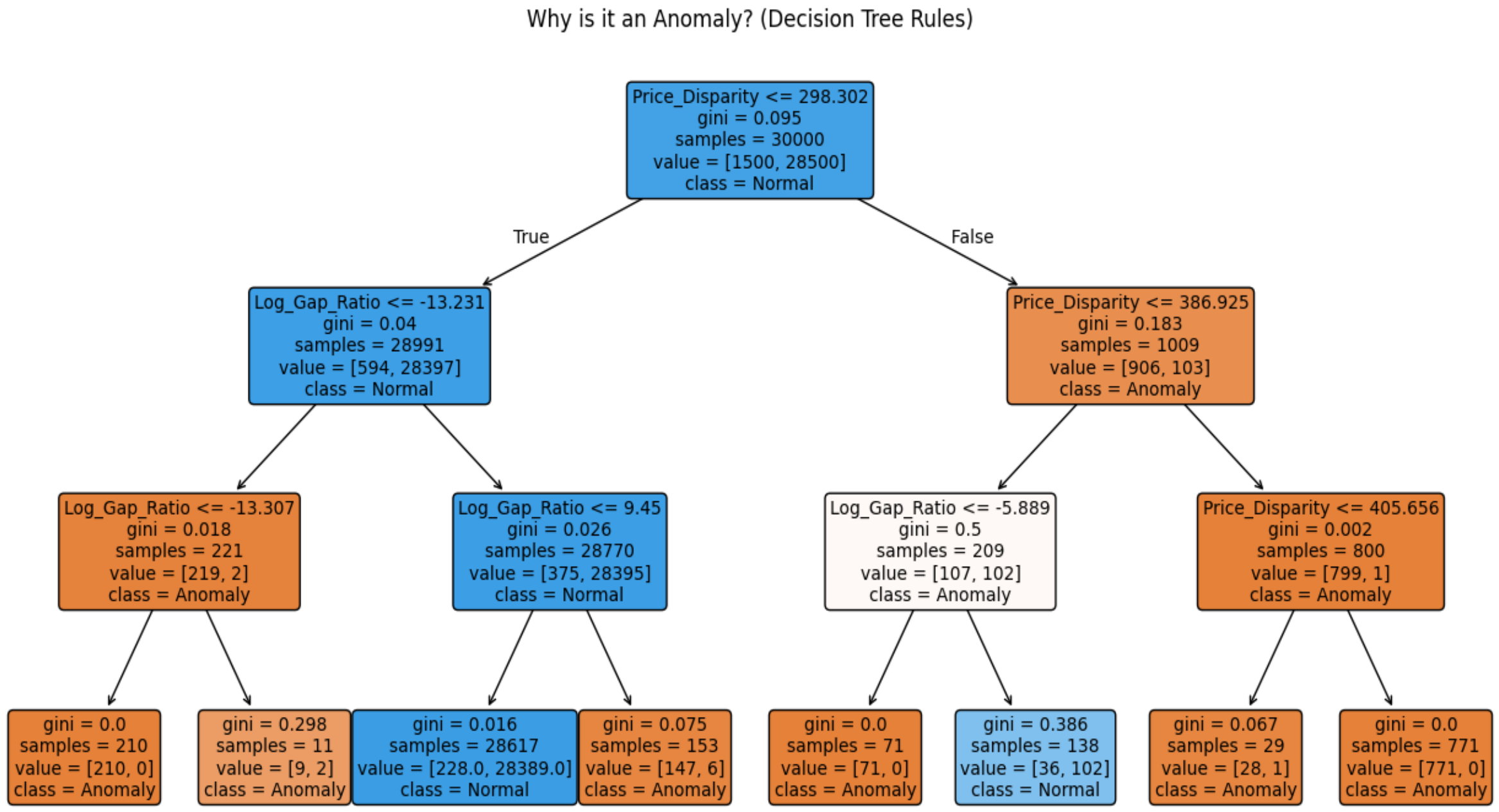

Source: Processed by Author (2025)

### 3.3 Structural Network Analysis (Layer 3)

To detect 'Geopolitical Rerouting,' the third layer employs Network Science to map the topology of global trade. Drawing on Vidmer et al.'s (2015) application of complex systems theory to international trade, we calculate centrality metrics to identify 'bridge' nodes that connect otherwise disparate trading communities. Furthermore, we utilize the modularity optimization techniques described by Newman (2006) to algorithmically detect 'Global Trade Clans,' identifying anomalies that manifest as structural violations of established trading communities.

**Figure 17. Network Graph Identifying High-Centrality Transshipment Hubs.**

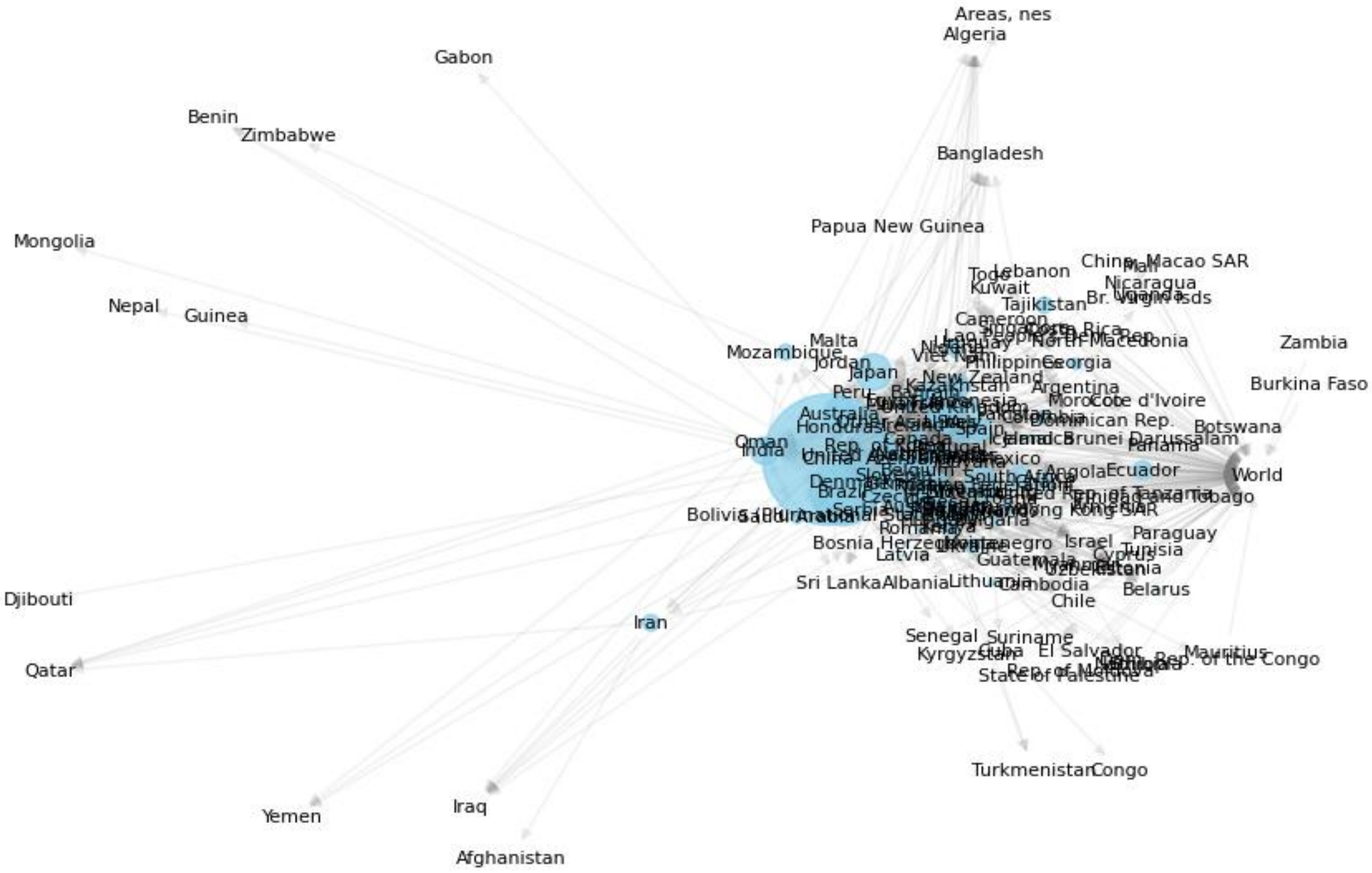

Source: Processed by Author (2025)

Figure 17 visualizes these high-centrality transshipment hubs, physically locating intermediaries (the large blue nodes) that facilitate indirect trade flows. Complementing this, the "Global Trade Clans" visualization in Figure 18 uses algorithmic community detection to color-coded distinct trading blocs. This analysis reveals that global trade is highly structured; therefore, anomalies often manifest as "community violations", trades that break established partnership patterns, such as a country suddenly trading heavily outside its algorithmic "clan".

**Figure 18. Algorithmic Community Detection of Global Trade Clans and Partnerships.**

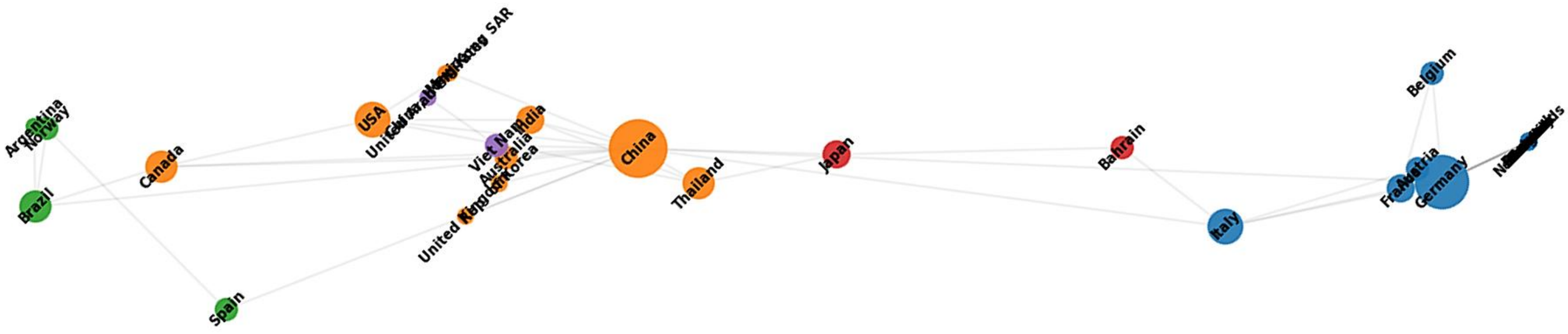

Source: Processed by Author (2025)

### 3.4 Non-Linear Value Analysis (Layer 4)

The final layer addresses complex, non-linear outliers using Deep Autoencoders. As reviewed by Chalapathy and Chawla (2019) and applied recently by Drammeh (2025), autoencoders compress data and attempt to reconstruct it; high reconstruction errors signal anomalies.

**Figure 19: UMAP Projection Visualizing the High-Risk Clusters.**

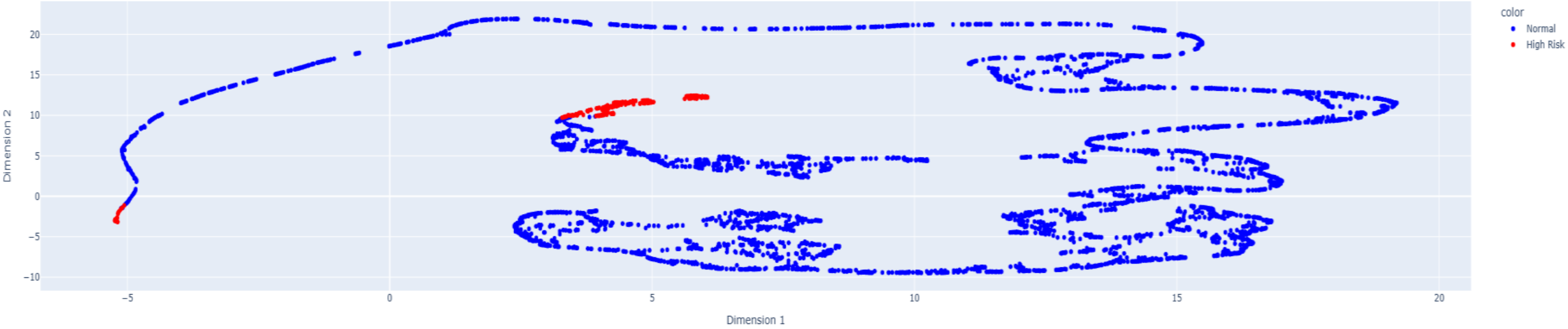

Source: Processed by Author (2025)

The "Reconstruction Error Analysis" in Figure 20 plots these error scores. The vast majority of normal trades (blue dots) sit near zero error, while the anomalies (red dots) spike upward, validating that the model successfully isolates complex outliers that linear models might miss. Finally, the "UMAP Projection" in Figure 19 uses Uniform Manifold Approximation and Projection, as defined by McInnes et al. (2018), to visualize these high-dimensional findings in 2D space. This projection reveals a distinct "High Risk" cluster (red) separated from the main "Normal" trade manifold (blue), effectively displaying an 'Island of Anomalies' that warrants immediate investigation.

**Figure 20. Deep Autoencoder Reconstruction Error Analysis for Non-Linear Anomalies.**

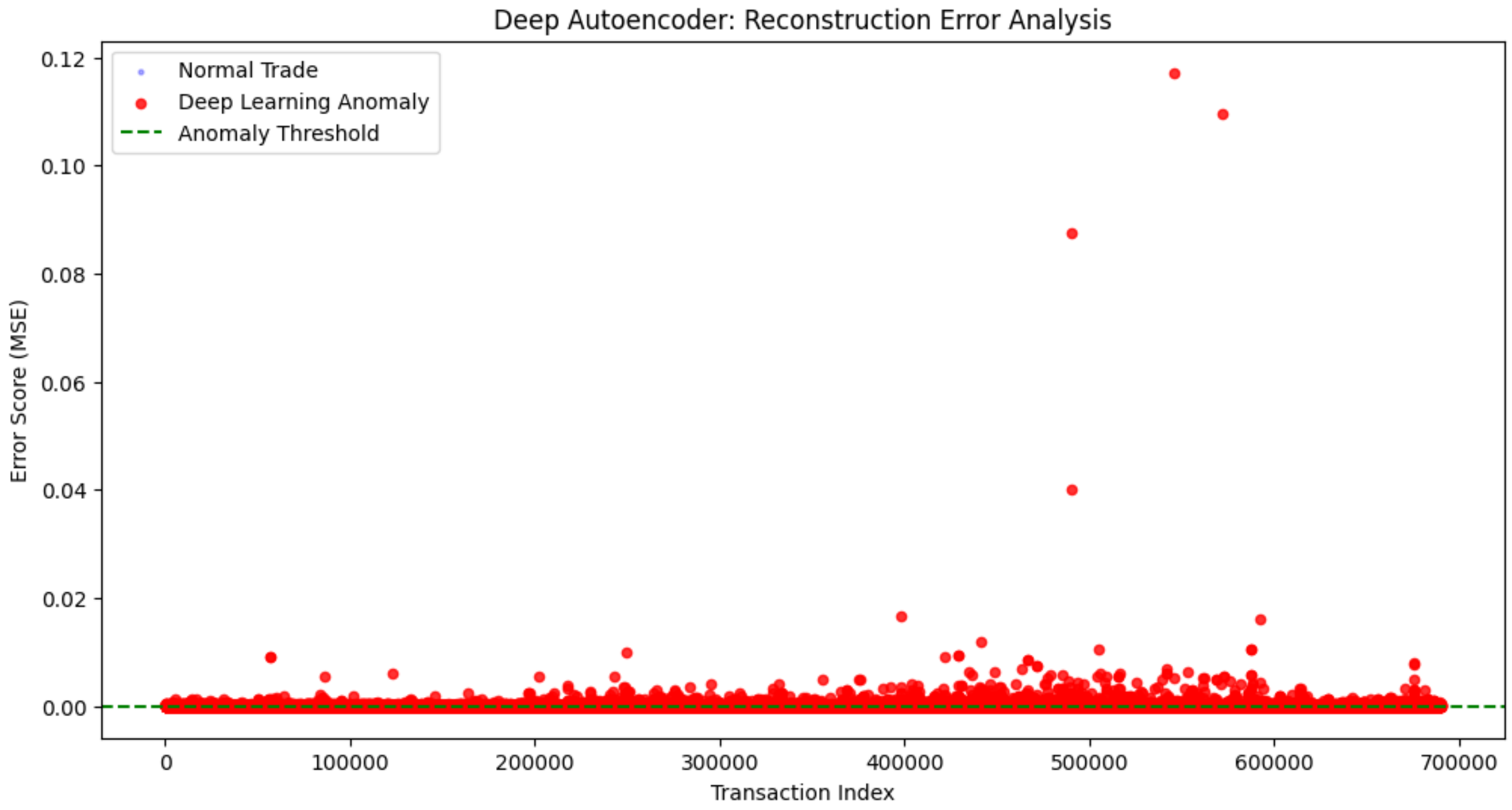

Source: Processed by Author (2025)

## 4. Empirical Results

### 4.0 Anatomy of Global Trade Anomalies

The application of the multi-layered pipeline revealed three distinct categories of systemic irregularities: large-scale volume gaps in intra-regional trade, complex value mismatches indicative of money laundering rather than simple greenwashing, and the specific routing of illicit flows through intermediary hubs.

### 4.1 Volume Discrepancies in Intra-Regional Trade

Analysis of "Mirror Statistics" provided the primary evidence for large-scale volume manipulation. Following the methodology established by Carrère and Grigoriou (2014) for detecting trade integration discrepancies, and the World Customs Organization’s framework for mirror analysis described by Cantens (2015), we compared reported exports versus imports for bilateral pairs. This dual-validation approach reveals massive deviations from the theoretical "Perfect Symmetry" line.

**Figure 21. Scatter Plot of Global Reported Export vs. Import Discrepancies (Log Scale).**

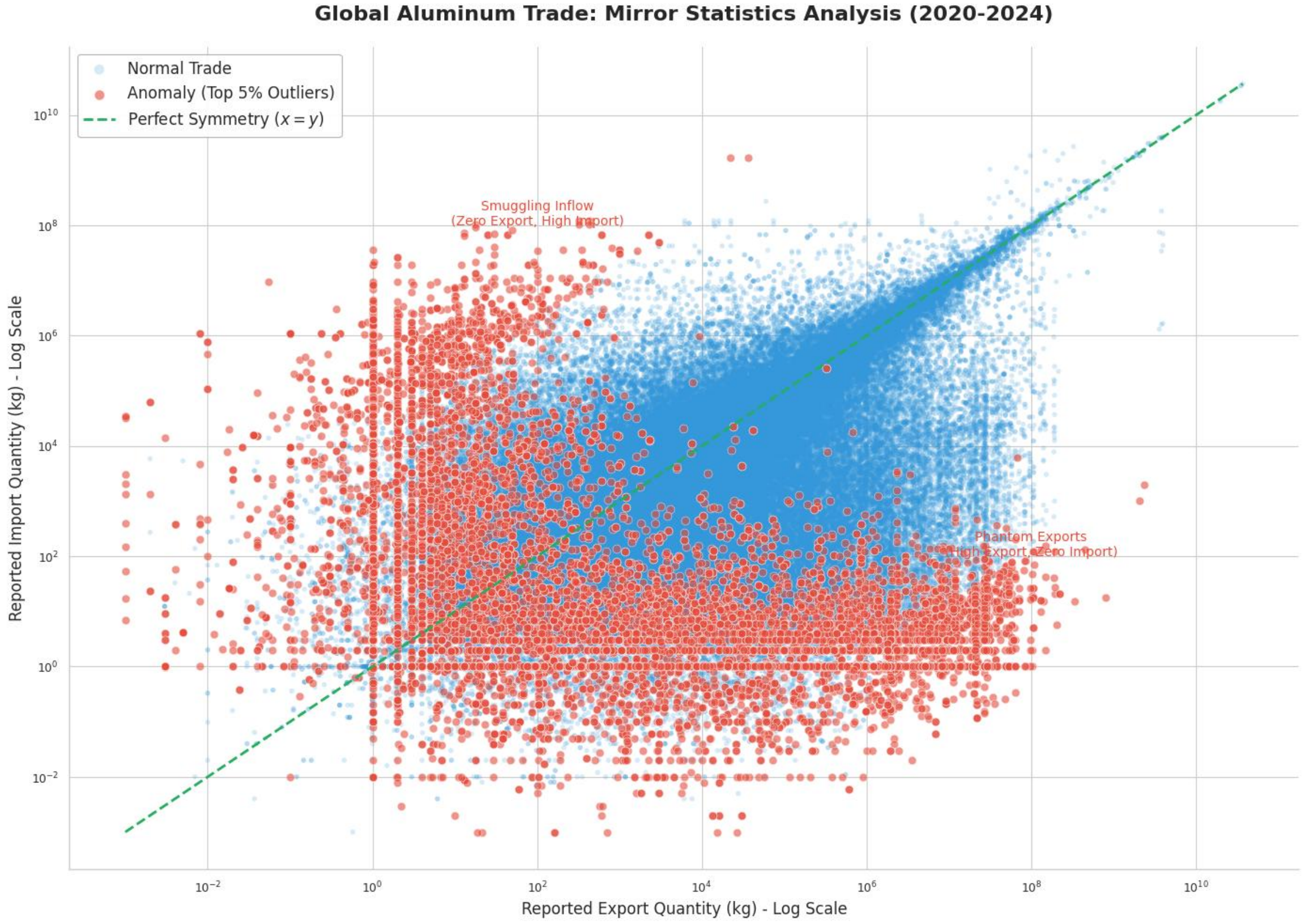

Source: Processed by Author (2025)

As visualized in Figure 21 above, the scatter plot identifies distinct "Phantom Exports", where nations report exporting millions of tons that never arrive, and "Smuggling Inflows," where imports appear without corresponding export declarations. The chart plots a green dashed line representing a perfect match; the red dots represent the "Top 5% Outliers" that deviate drastically from this line, constituting the bulk of the "Missing Metal" problem. These discrepancies are further contextualized by analyzing product-specific behaviors. Figure 22 demonstrates that these anomalies are not uniform but product-specific. By comparing the boxplots of Scrap (HS 7602) against Semi-Finished goods (HS 7604/7606), we observe significant overlap and outliers. This visual "sanity check" suggests that fraud follows specific "physics" depending on the industrial application; for instance, semi-finished goods are occasionally trading at suspiciously low prices that overlap with scrap values, indicating potential misclassification. This is reinforced by the line chart on the right of Figure 22, which shows that as global trade volume (the red line) declines, the remaining trade becomes statistically riskier.

**Figure 22. Unit Price Comparison of Scrap vs. Semi-Finished Goods and Global Volume Trends.**

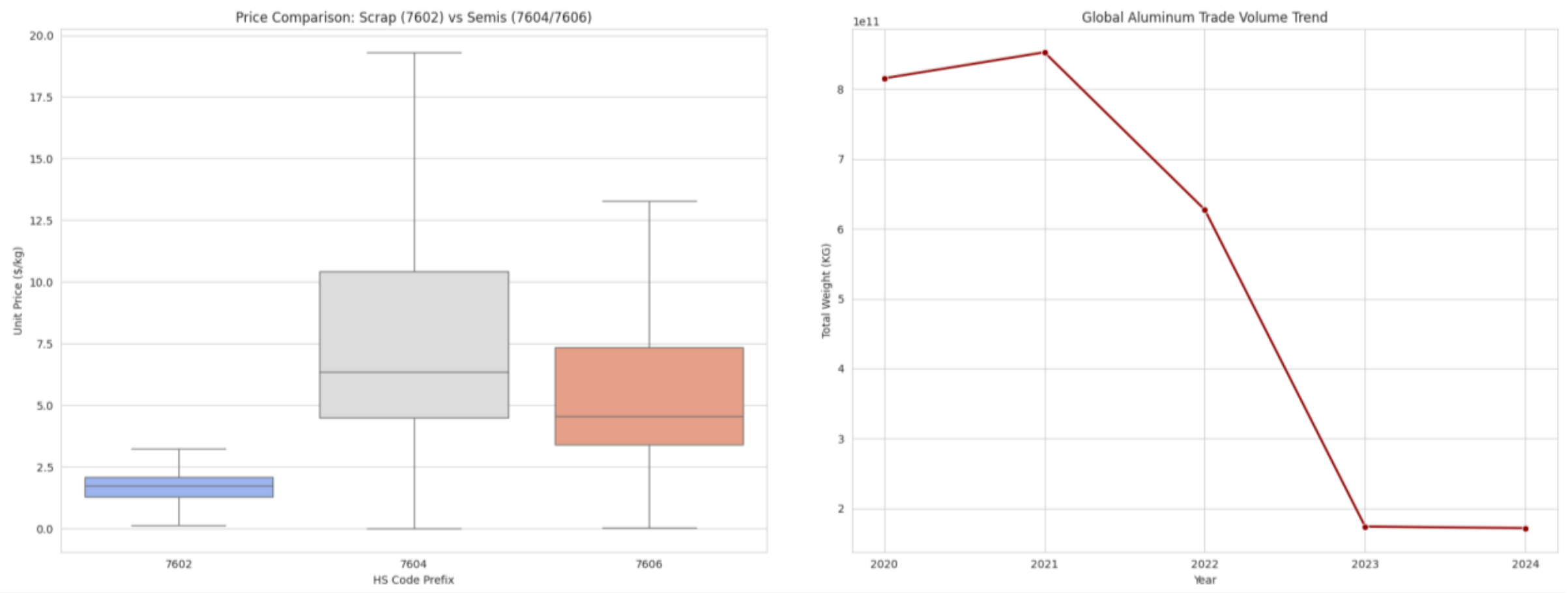

Source: Processed by Author (2025)

**Figure 23. Multi-Tiered Scatter Plot Analysis of Price vs. Volume across HS Codes.**

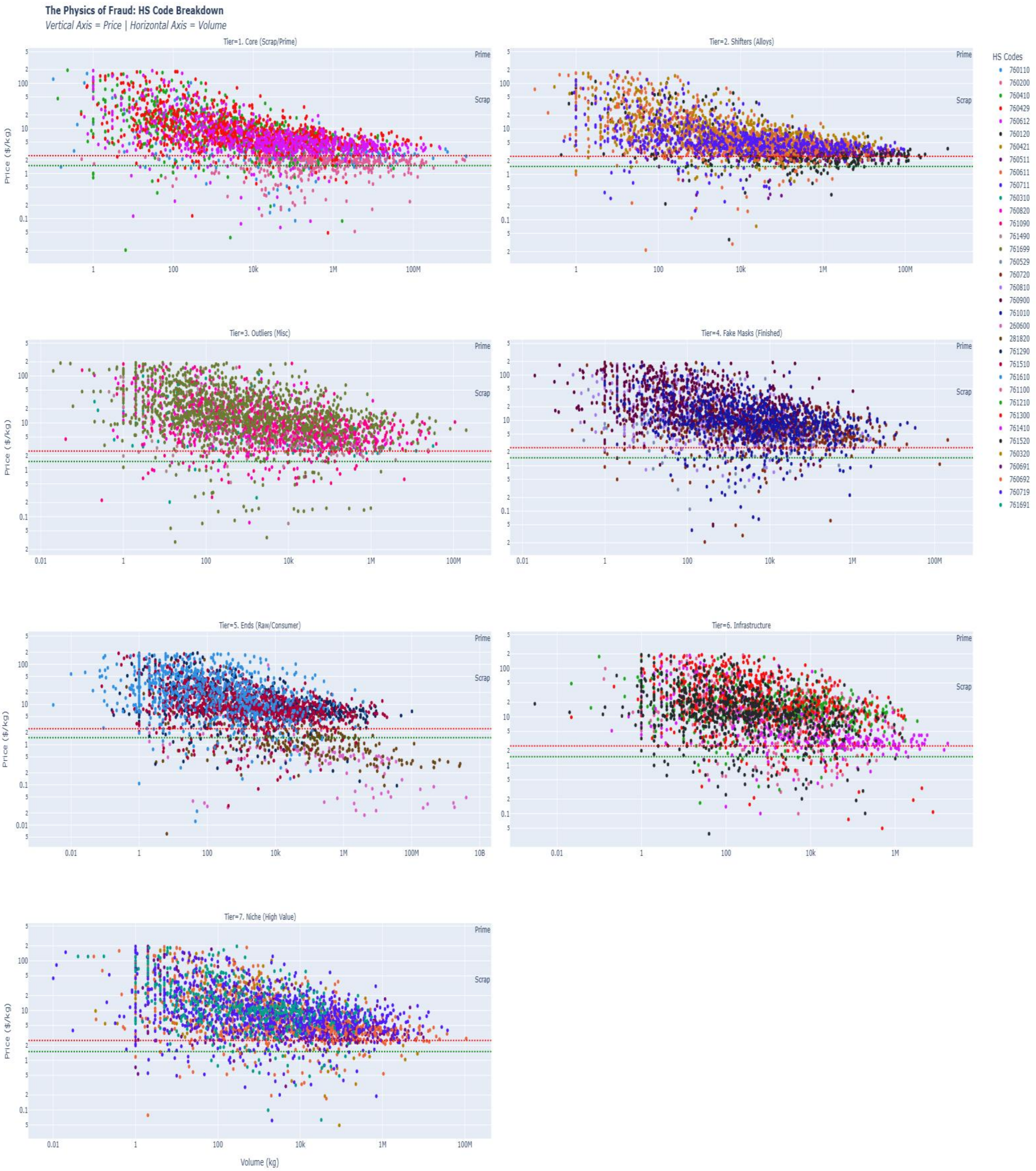

Source: Processed by Author (2025)

Deepening this analysis, Figure 23 breaks down these anomalies into specific tiers based on the Harmonized System (HS) codes. The data demonstrates that "Alloy" anomalies (Tier 2) exhibit distinct price and volume clustering compared to "Waste" anomalies (Tier 1). This distinction suggests that fraud is not a monolith but follows specific patterns depending on the intrinsic value of the metal involved, requiring different detection thresholds for raw waste versus processed alloys.

### 4.2. Value Mismatches and Pricing Anomalies

To establish a baseline for pricing irregularities, we first analyzed the frequency distribution of global aluminium export prices. The histogram below overlays a normal distribution curve and a clear red dashed line marking the global median price of $9.16/kg.

**Figure 24. Histogram of Global Aluminium Export Prices with Median Thresholds.**

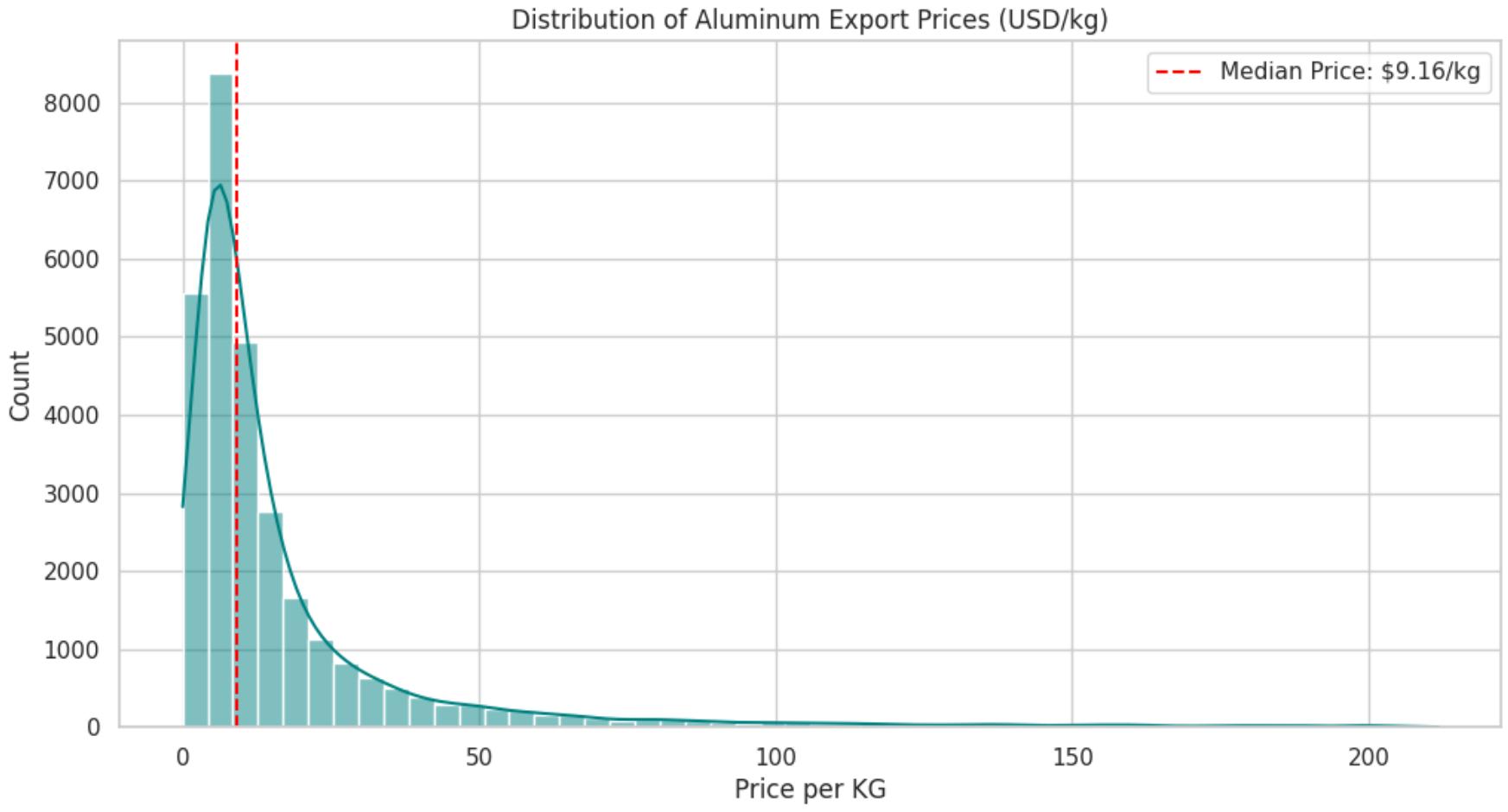

Source: Processed by Author (2025)

The primary purpose of Figure 24 is to visualize the "Long Tail" of price anomalies, the transactions that fall far to the right of the curve at extreme price points exceeding $100/kg. By establishing the statistical "center" of global pricing, this figure provides the mathematical baseline used to flag "Price Deviation" anomalies. Building on this baseline, the unsupervised Isolation Forest algorithm successfully isolated a cluster of high-risk transactions characterized by extreme unit-price deviations.

**Figure 25. Evidence of Trade-Based Money Laundering: The "Price Physics Breach".**

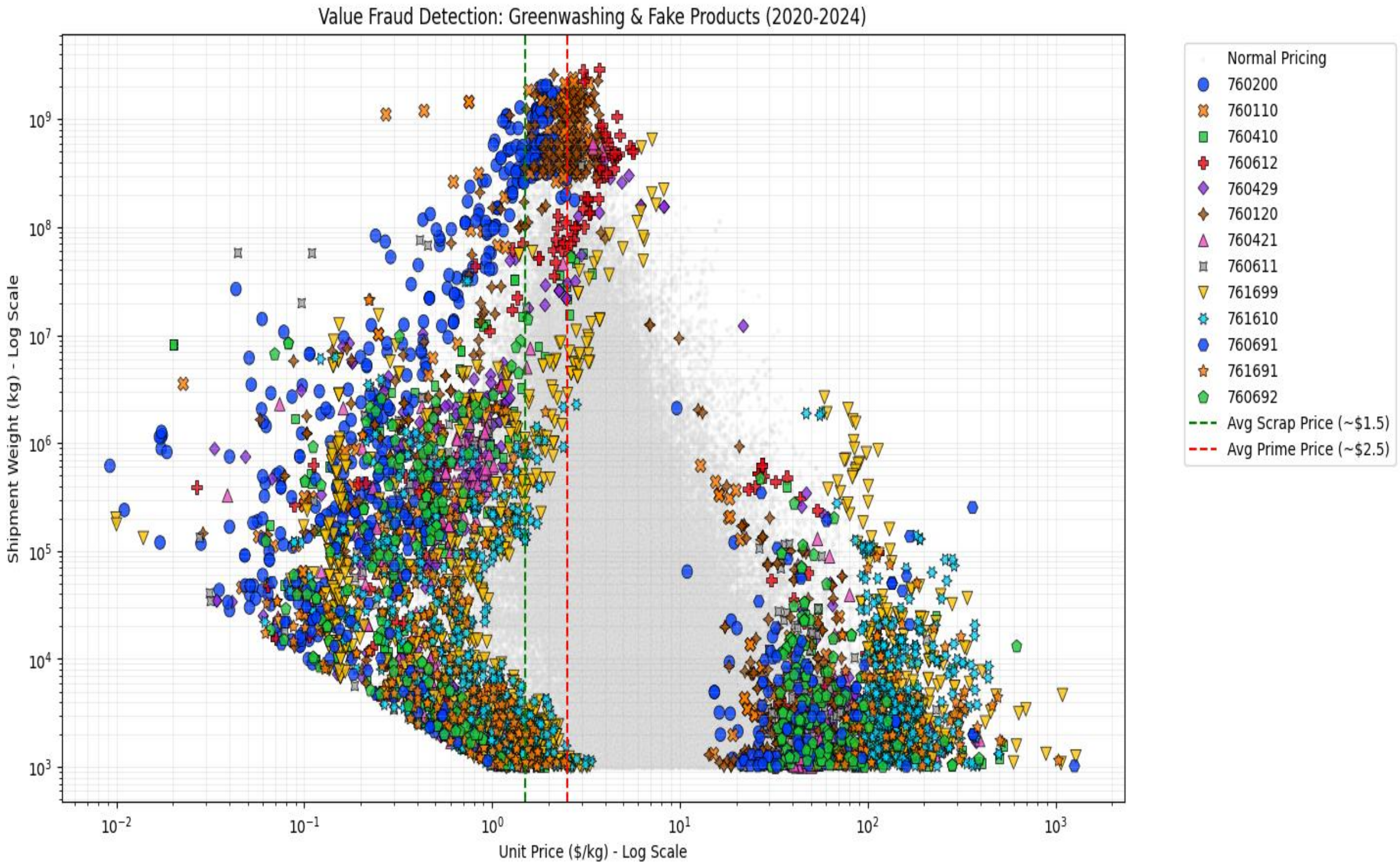

Source: Processed by Author (2025)

While the global median price for aluminium is approximately $8.23/kg, the algorithm identified a distinct high-risk cluster trading at a median of $167.59/kg, a markup of over 1,900%. This is visually confirmed by Figure 25, the "Price Physics Breach" chart. This box plot highlights the statistical separation between "Normal" trade (in blue) and the "High Risk" cluster (in red), which sits far to the right of the value distribution. This extreme over-valuation contradicts the "Greenwashing" hypothesis and instead points to systematic TBML via over-invoicing.

**Figure 26. Frequency Distribution of High-Risk Harmonized System (HS) Codes (The "Hardware Mask").**

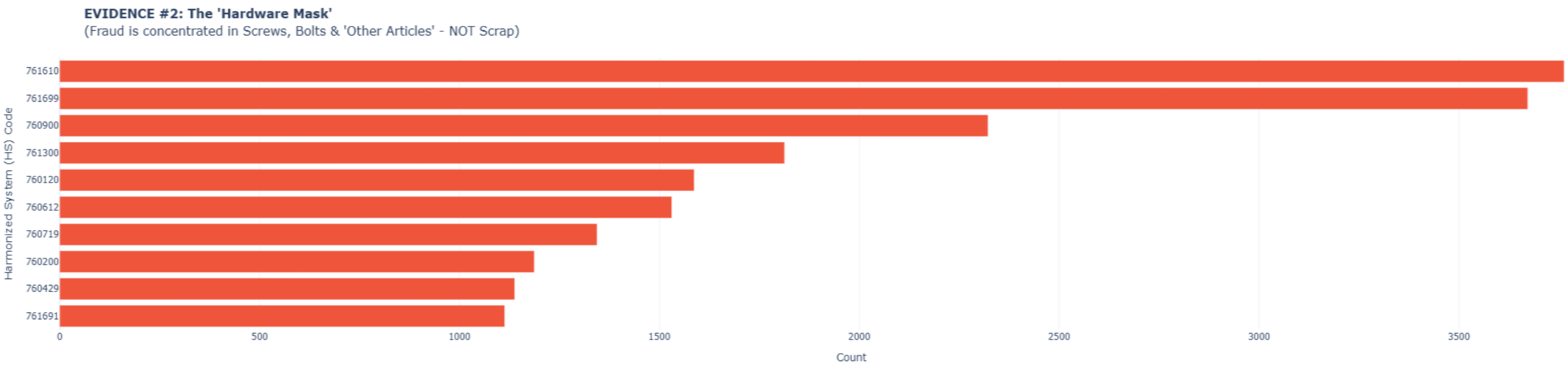

Source: Processed by Author (2025)

Contrary to the initial expectation that fraud would concentrate in raw scrap, the analysis revealed a distinct "Hardware Masking" strategy centered on heterogeneous categories like HS 7616 (Nails, Screws, Bolts), which account for over 21% of anomalies. This behavior represents a calculated "fiscal play" to exploit the conflicting incentives revealed in Figure 2 and Figure 3: while the import landscape incentivizes declaring goods as 0% duty "Scrap" to avoid high tariffs on finished articles, the export landscape systematically inverts this logic, imposing a punitive 10% duty on "Waste & Scrap" while leaving finished goods duty-free. Consequently, illicit actors are compelled to execute a bi-directional misclassification strategy, exporting restricted scrap disguised as duty-free "hardware" to evade exit levies, then re-declaring the same cargo as duty-free "scrap" upon arrival to evade entry taxes. This strategic code-switching is the primary engine of the "Mirror Fracture," as the resulting mismatch between the export record (Hardware) and the import record (Scrap) prevents bilateral reconciliation, creating the statistical illusion that massive volumes of metal are vanishing from the global ledger.

**Figure 27. Longitudinal Shifts in East Asian Export Destinations and Transshipment Routes.**

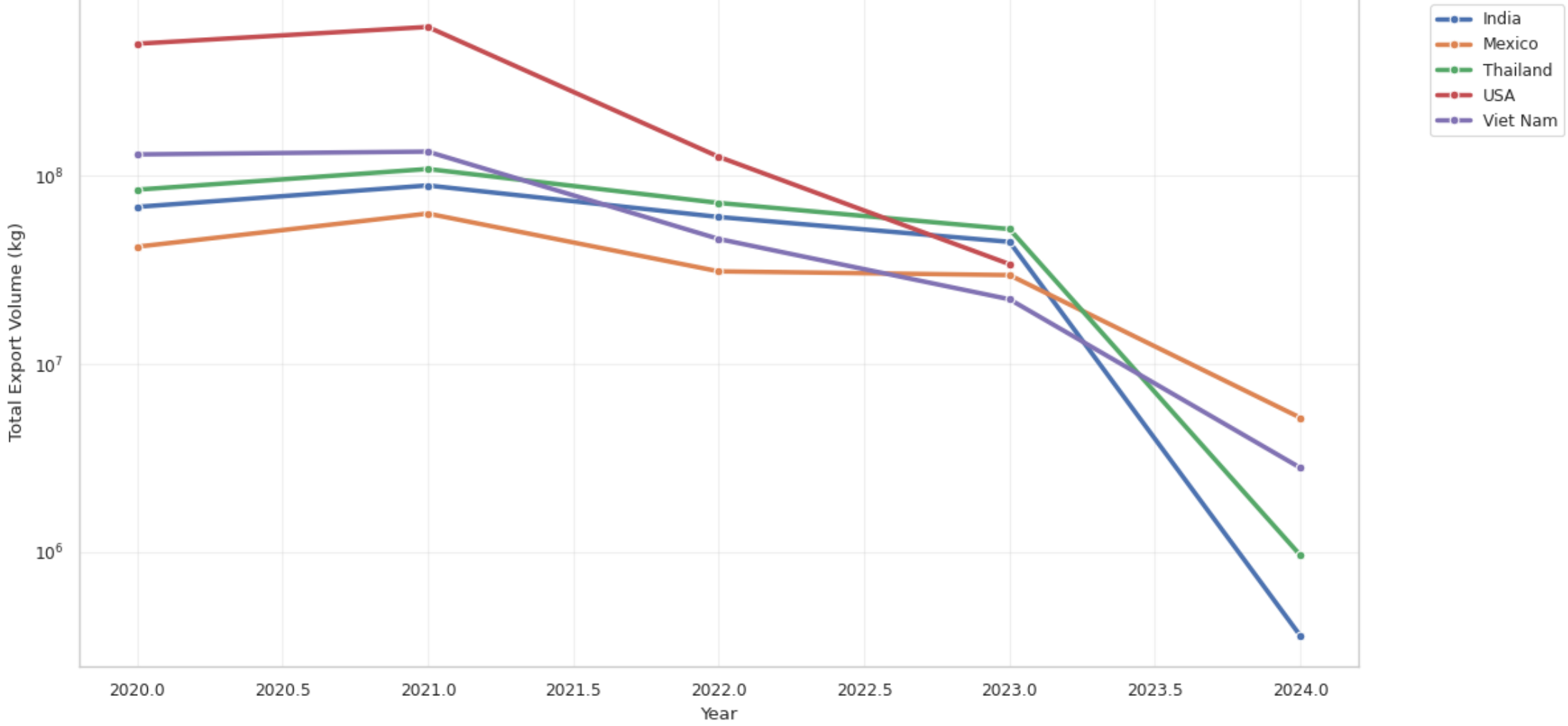

Source: Processed by Author (2025)

### 4.3. Identification of Intermediary Hubs and "Ghost Routes"

Furthermore, the network analysis exposed a sophisticated evasion dynamic where trade flows do not merely shift geographically but disappear statistically. We identify this strategic maneuver as "Void-Shoring": a technique where exporters actively route high-value assets into a data void rather than a physical jurisdiction. Unlike traditional offshoring, which seeks favorable labor or tax conditions in a specific country, Void-Shoring directs flows to "Unspecified" (Code 0) destinations, effectively severing the forensic link between origin and final consumption. This is evidenced by the "Ghost Destination" pattern, where major exporters such as the United Kingdom and Norway were found routing high-value anomalies exclusively to these undefined locations. By utilizing Void-Shoring, these actors successfully shield their transactions from mirror analysis, ensuring that while the physical metal moves through the supply chain, its digital twin vanishes from the regulatory map. Figure 27 tracks the rapid displacement of export volumes from East Asian manufacturing hubs between 2020 and 2024. As direct exports to major North American consumer markets declined, trade volumes concurrently surged in alternative intermediary hubs across South Asia and Latin America. A critical finding within these shifts is the "Ghost Destination" pattern, which statistically resembles the 'quasi-transit' phenomenon described by Lemmers and Wong (2019).

**Figure 28. The "Ghost Route" Mechanism: Systematic Destination Suppression.**

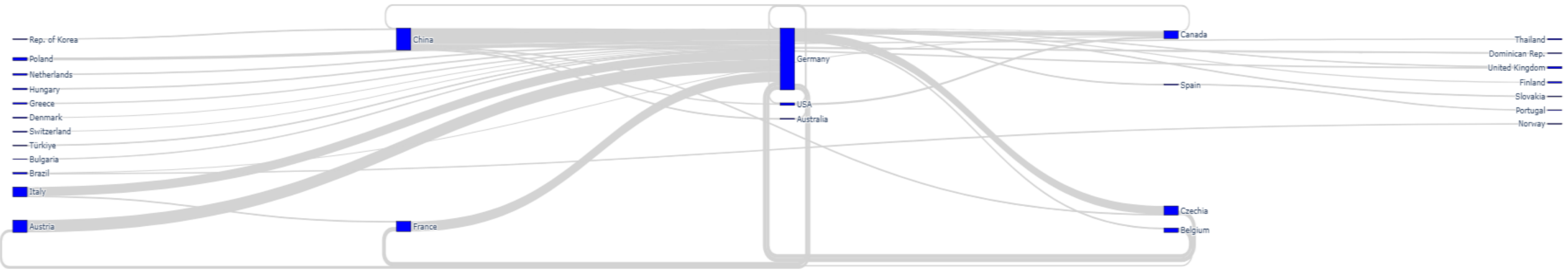

Source: Processed by Author (2025)

Figure 28 illustrates that the top fraud corridors, originating from major exporters in Western Europe and Southeast Asia, predominantly list their destination as 'Unspecified' (Code 0), represented by the massive top bar in the chart. This systemic suppression of destination data effectively fractures the 'mirror statistics' mechanism, preventing reconciliation and distinguishing genuine re-exports from illicit diversion. Most notably, high-value anomalies originating from key Northern and Western European markets (reaching valuations of $323/kg and $384/kg) were routed exclusively to these undefined destinations.

**Figure 29. Predicted True Origins of Intermediary Shipments from Transshipment Hubs.**

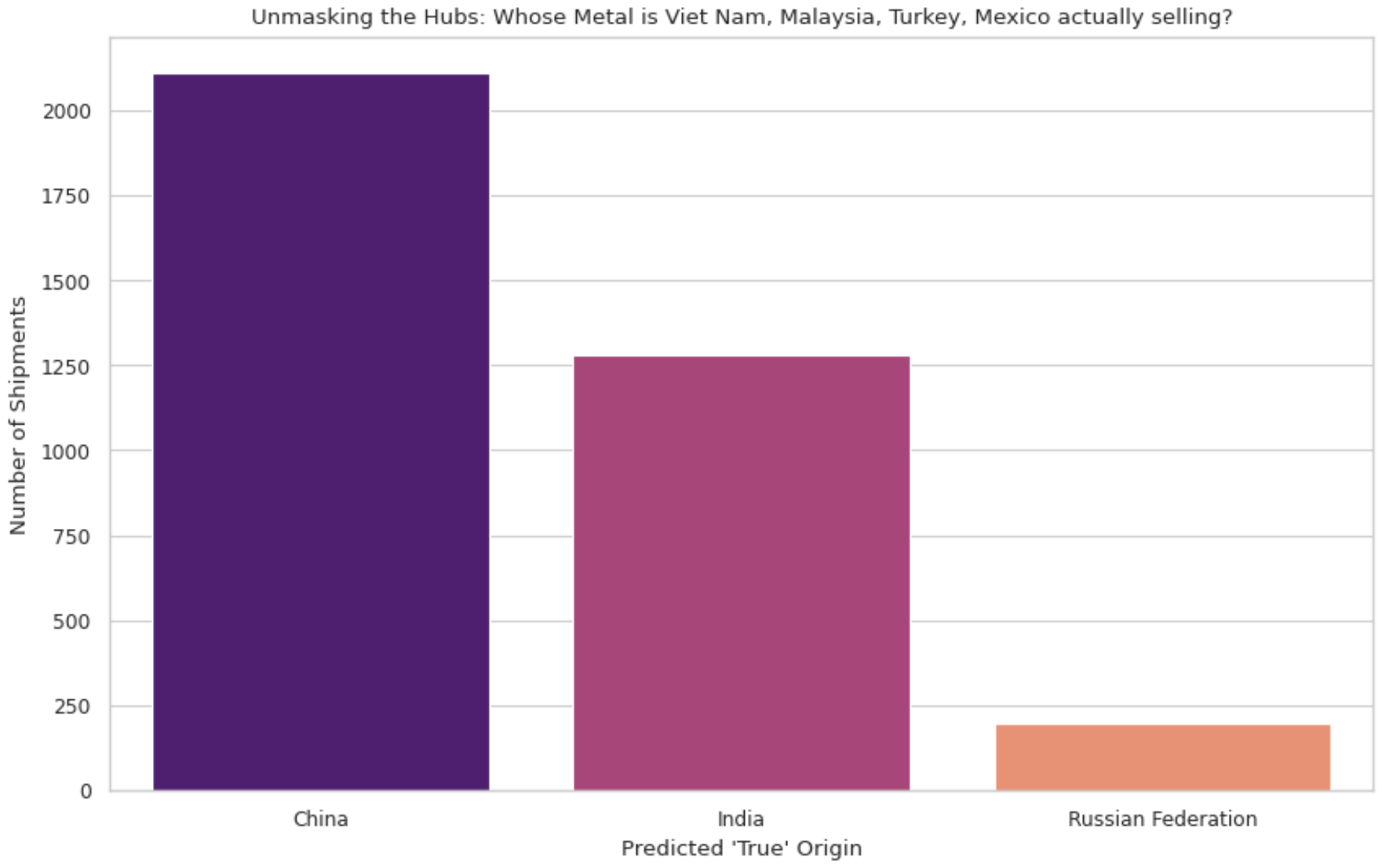

Source: Processed by Author (2025)

Further unmasking these flows, Figure 29 uses predictive modeling to estimate the true origin of shipments passing through intermediary hubs. The data estimates that a massive portion of shipments from intermediaries in Southeast Asia and the Near East actually originate from East Asia (the large purple bar). This quantifies the volume of metal being re-labelled to obscure its true source, confirming the "Geopolitical Rerouting" typology and validating the hypothesis that transshipment hubs are being used to bypass origin-based tariffs.

### 4.4 Geospatial Analysis of Risk

Finally, the "Global Risk Map" utilizes heatmaps to visualize the geographic concentration of suspicious volumes. Rather than being randomly distributed, the map highlights that anomalies are heavily concentrated in specific regions. The color intensity on the map in Figure 30 reveals that while the Global North often acts as the financial destination for value transfer, the physical anomalies are frequently clustered around specific transshipment hubs in the Global South and East Asia. This static view is complemented by dynamic analysis to understand the evolution of these corridors.

**Figure 30. Geospatial Heatmap of Suspicious Aluminium Export Volumes.**

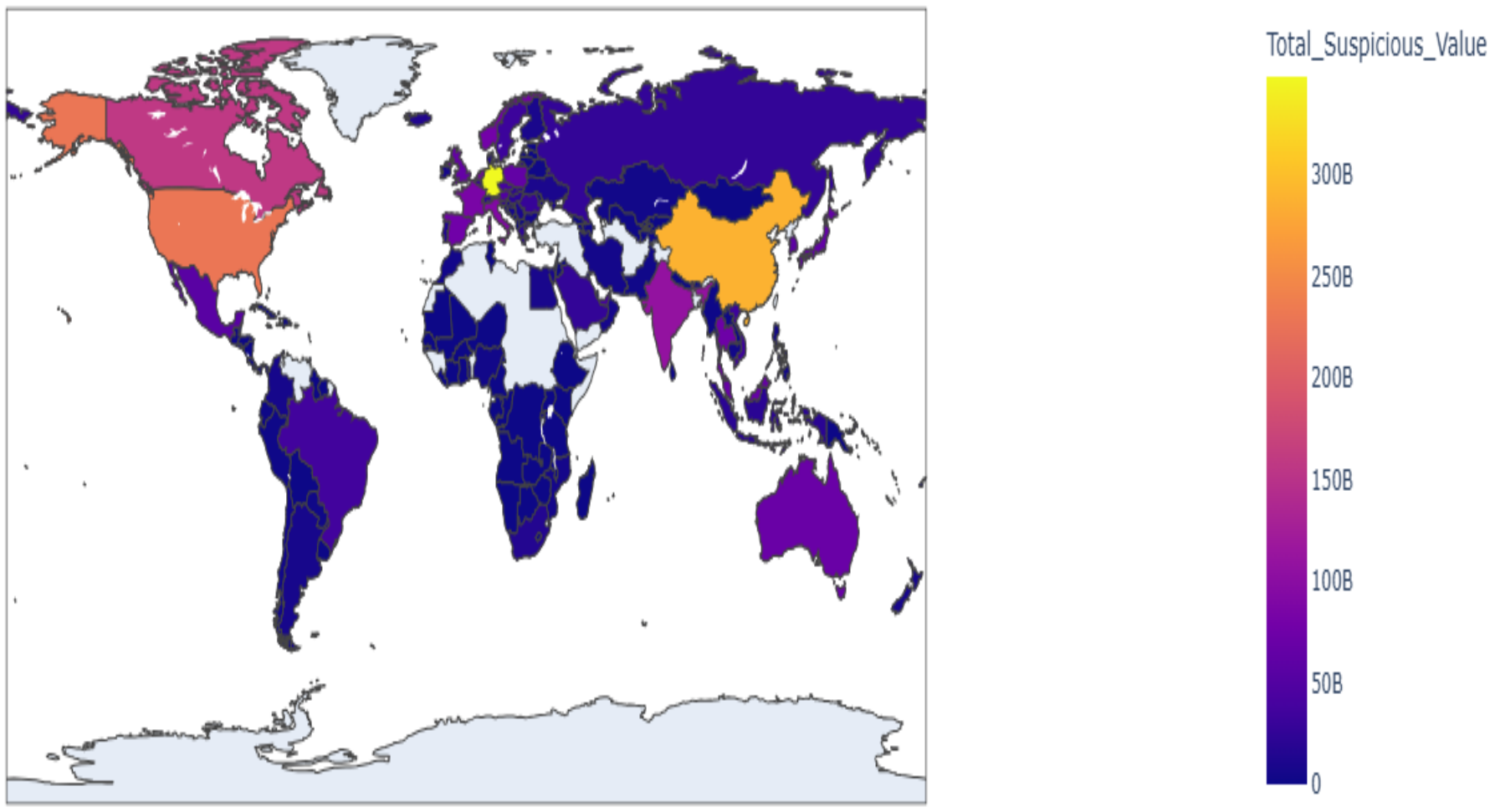

Source: Processed by Author (2025)

**Figure 31. Geospatial Analysis of Illicit Trade Corridors and Data Suppression**

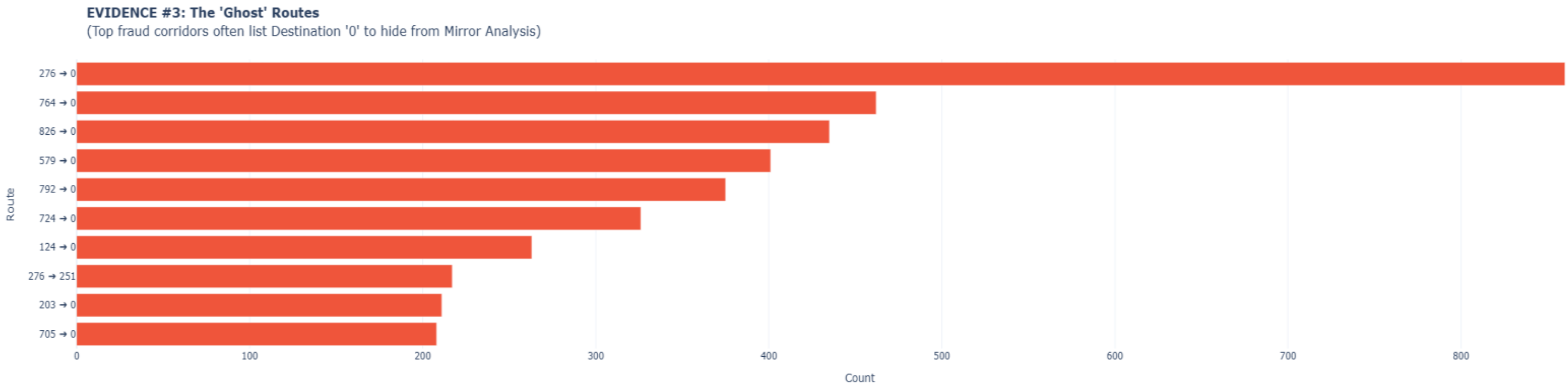

Source: Processed by Author (2025)

The analysis identifies a structural evolution in trade evasion techniques, revealing that illicit actors have moved beyond simple concealment to a strategic suppression maneuver termed "Void-Shoring," which actively routes high-value assets into a statistical "void" rather than a physical jurisdiction. The findings demonstrate that major exporters in Western Europe and Southeast Asia are systematically declaring destinations as "Unspecified" (Code 0) to sever the forensic link between origin and final consumption, effectively neutralizing mirror statistics analysis. This "Ghost Destination" mechanism is heavily correlated with extreme financial anomalies; notably, the study detected that high-value exports originating from key European markets were routed exclusively to these undefined locations, suggesting that the most severe price deviations rely on this data suppression to evade regulatory visibility.

The dynamic analysis in Figure 32 animates these routes over time, showing how the arcs of illicit trade physically migrate to new pathways in response to regulatory pressure. This underscores the need for dynamic rather than static monitoring systems to track the evolving geography of trade fraud. Geospatially, the research confirms the existence of a highly reactive and compartmentalized fraud network, where transshipment hubs are used to mask the true geopolitical origin of goods. Predictive modeling unmasked that a substantial volume of shipments passing through intermediaries in Southeast Asia and the Near East actually originate from major East Asian producers, validating the "Geopolitical Rerouting" typology used to bypass origin-based tariffs. Furthermore, longitudinal analysis reveals that these illicit corridors are physically migratory; as direct exports from East Asia to major North American markets declined, trade volumes surged in alternative hubs within South Asia and Latin America, creating a distinct risk map where the Global North acts as the financial destination while the Global South serves as the physical conduit for evasion.

**Figure 32. Geospatial Visualization of Shifting Anomaly Routes Over Time (2020-2024).**

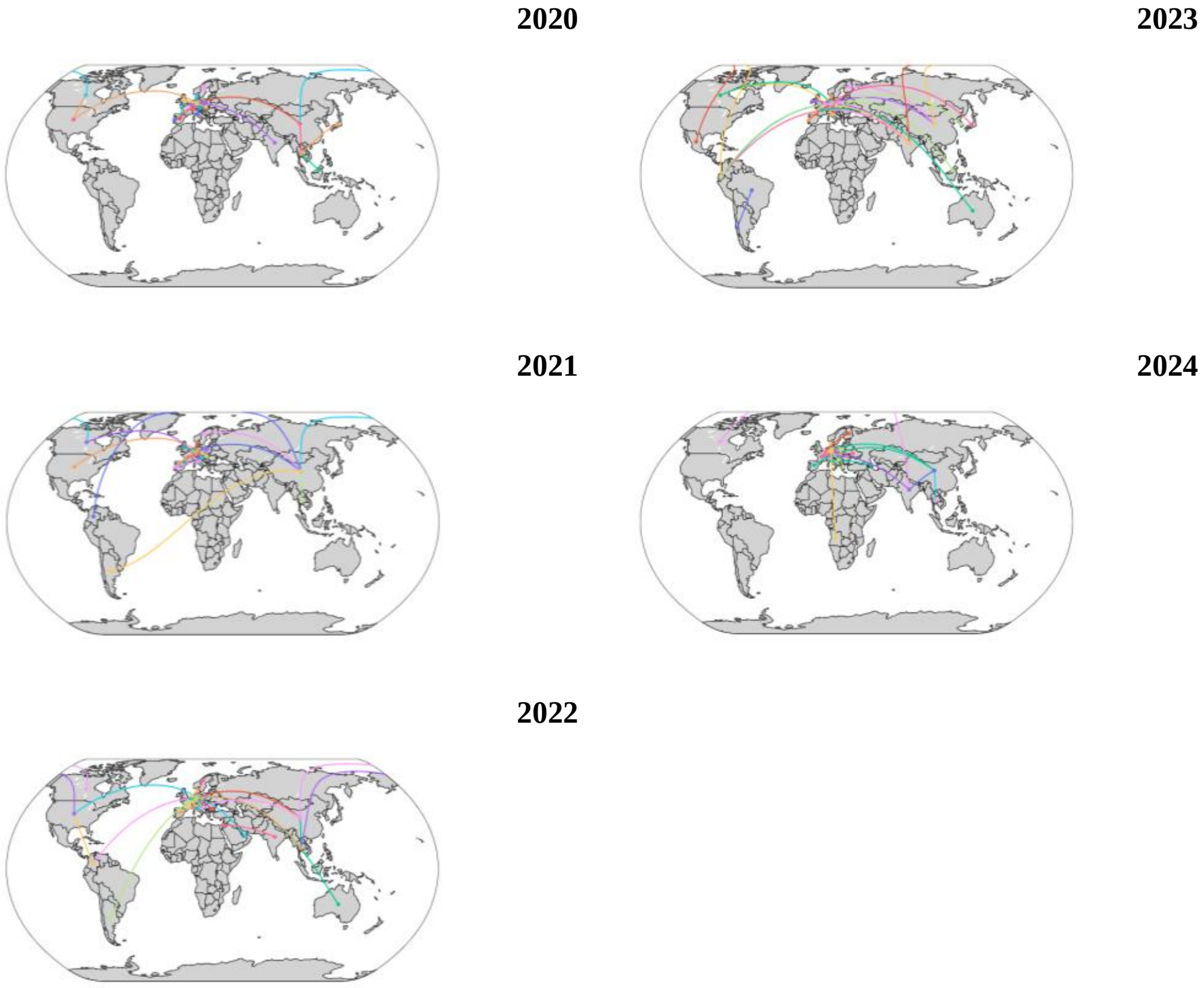

Source: Processed by Author (2025)

## 5. Discussion

### 5.0 Discussion: Drivers of Anomaly Detection

The empirical results confirm that trade fraud is not a random occurrence; rather, it possesses a distinct statistical signature driven by specific economic incentives. This discussion analyzes the specific features that drive algorithmic detection, the profiles of the actors involved, and the multivariate "fingerprint" that distinguishes high-risk trades from legitimate commerce.

### 5.1 Feature Importance (Explainable AI)

To decode the logic of the AI models and ensure transparency, we utilized SHAP (Shapley Additive Explanations), the unified framework for interpreting predictions proposed by Lundberg and Lee (2017). This approach addresses the "black box" problem highlighted by Molnar (2020) by assigning a specific contribution value to each feature for every prediction. As illustrated in Figure 33, the model identifies "Price_Deviation" and "Unit_Value" as the top predictors of fraud. The wide horizontal spread of SHAP values for price deviation confirms that it is the single most powerful signal, validating the hypothesis that price arbitrage, rather than simple volume, is the primary mechanism of illicit activity. This moves the detection capability from an opaque score to a transparent, actionable insight for regulators.

This relationship is further detailed in the Beeswarm Plot in Figure 34, which reveals the directional impact of these features. In this visualization, high values for "Price_Disparity" (represented by red dots) are strongly associated with anomaly classification, appearing far to the right of the baseline. In contrast, volumetric features like "Log_Gap_Ratio" show a more complex, bi-modal distribution. This distinction suggests that customs risk engines should weight pricing variables significantly higher than simple volumetric checks, as price anomalies are a more consistent indicator of fraud than volume mismatches.

**Figure 33. SHAP Summary Plot Identifying Key Drivers of Trade Anomalies.**

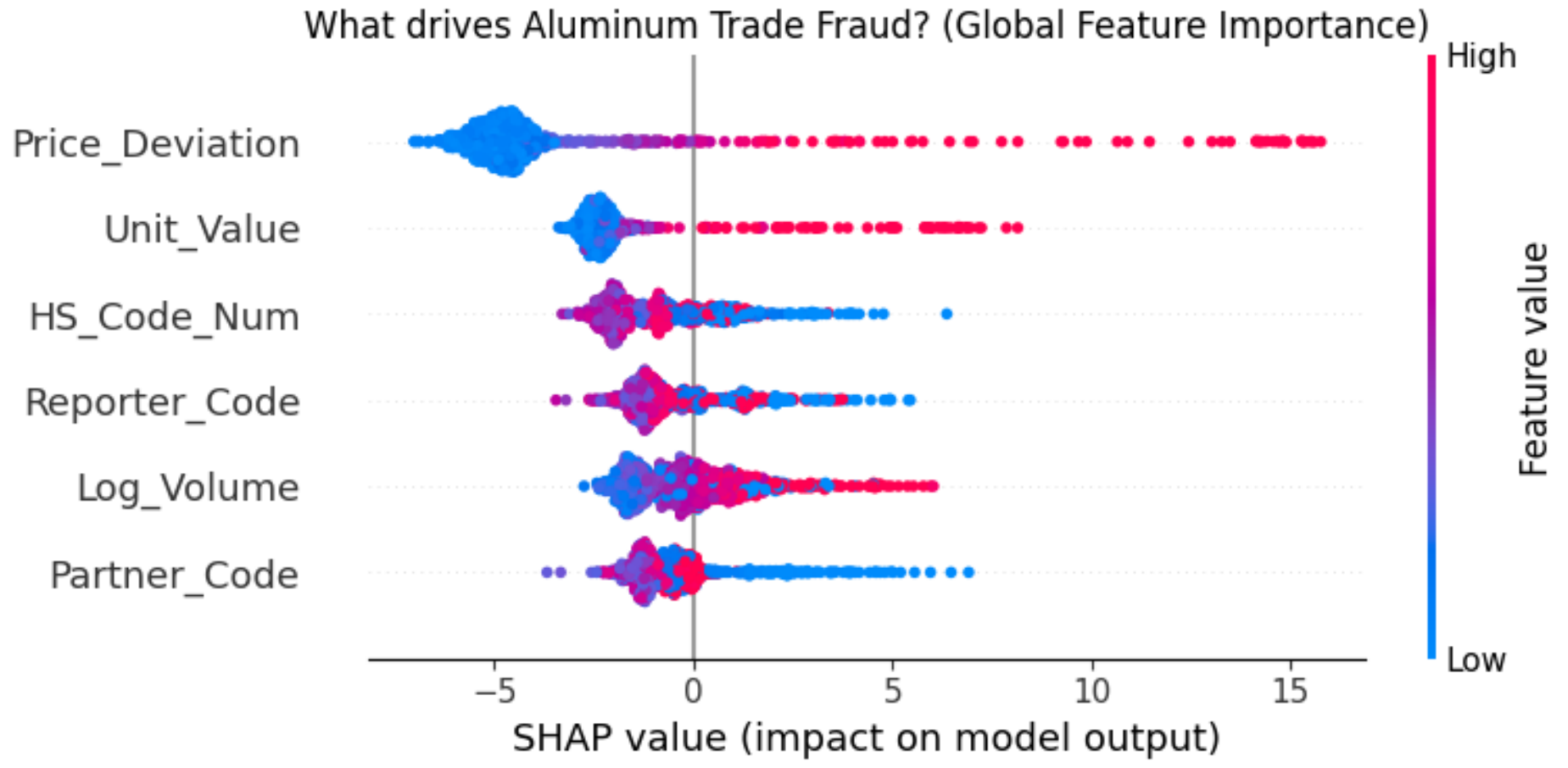

Source: Processed by Author (2025)

**Figure 34: SHAP Beeswarm Plot Analyzing Features Driving Anomaly Classification.**

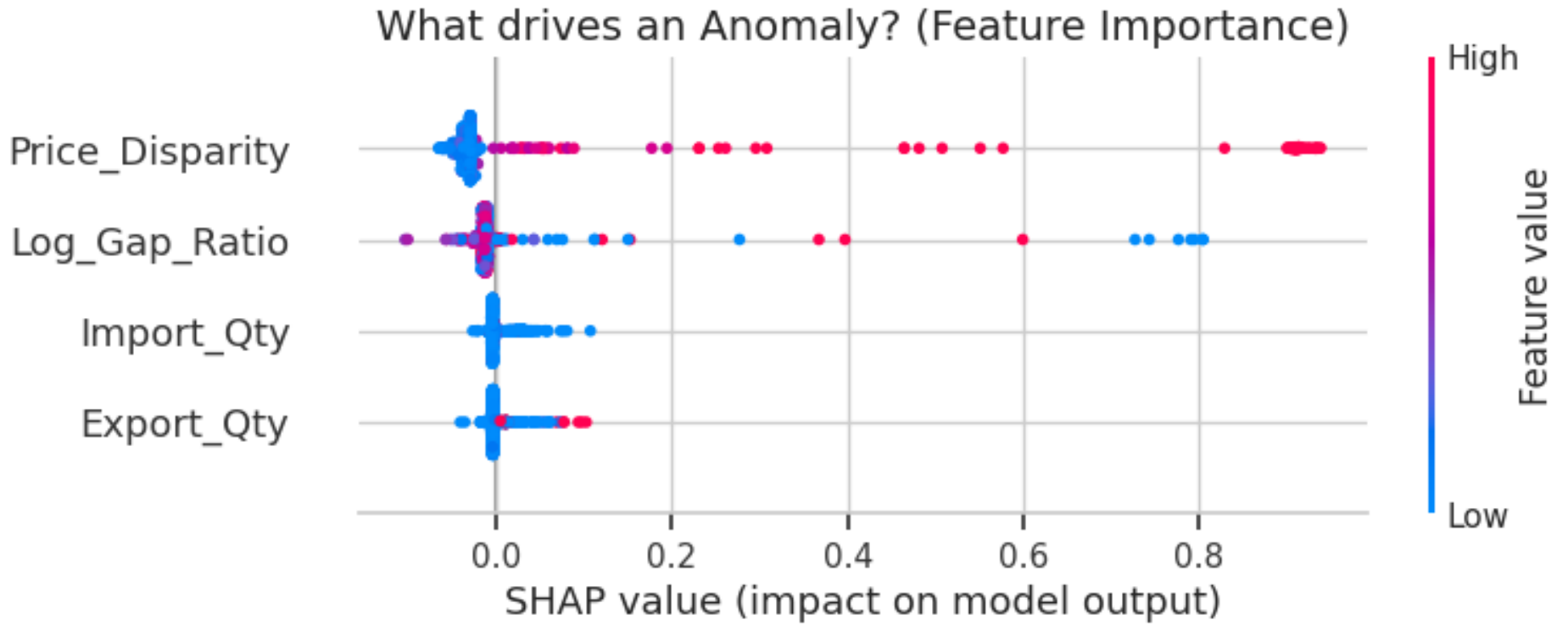

Source: Processed by Author (2025)

**Figure 35. Ranking of Feature Importance for Anomaly Detection using XGBoost.**

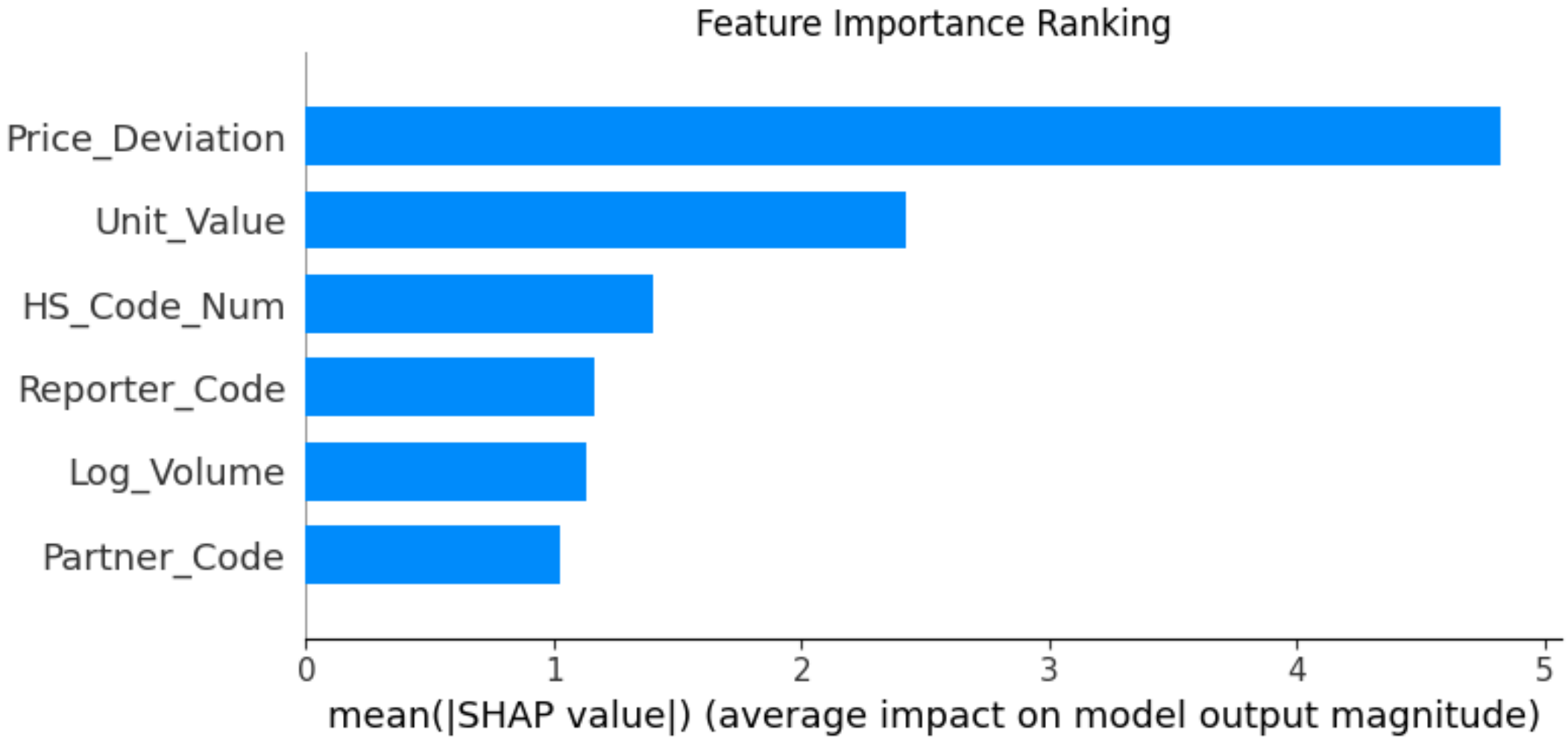

Source: Processed by Author (2025)

Complementing the SHAP analysis, Figure 35 provides a simplified, algorithmic ranking of feature importance specifically derived from the XGBoost classifier. It isolates the top six variables, showing a steep drop-off after "Price_Deviation" and "Unit_Value". This visual ranking serves as a clear technical blueprint for customs authorities, reinforcing the finding that while administrative factors like "Reporter_Code" (who is reporting) or "Log_Volume" (how much is traded) are relevant, they are secondary to the economics of the deal.

### 5.2 Actor Profiling and Market Segmentation

Moving from individual transactions to actor profiling, Figure 36 presents the results of K-Means clustering, which segregates reporting countries into distinct "Risk Profiles". In this scatter plot, high-volume, legitimate traders, cluster together in specific zones, while "Shadow Hubs", countries with unusual pricing structures but lower volumes are isolated in separate clusters for example the yellow and purple nodes. This clustering empirically separates standard market participants from high-risk actors.

**Figure 36. K-Means Clustering of Global Aluminium Actors by Risk Profile.**

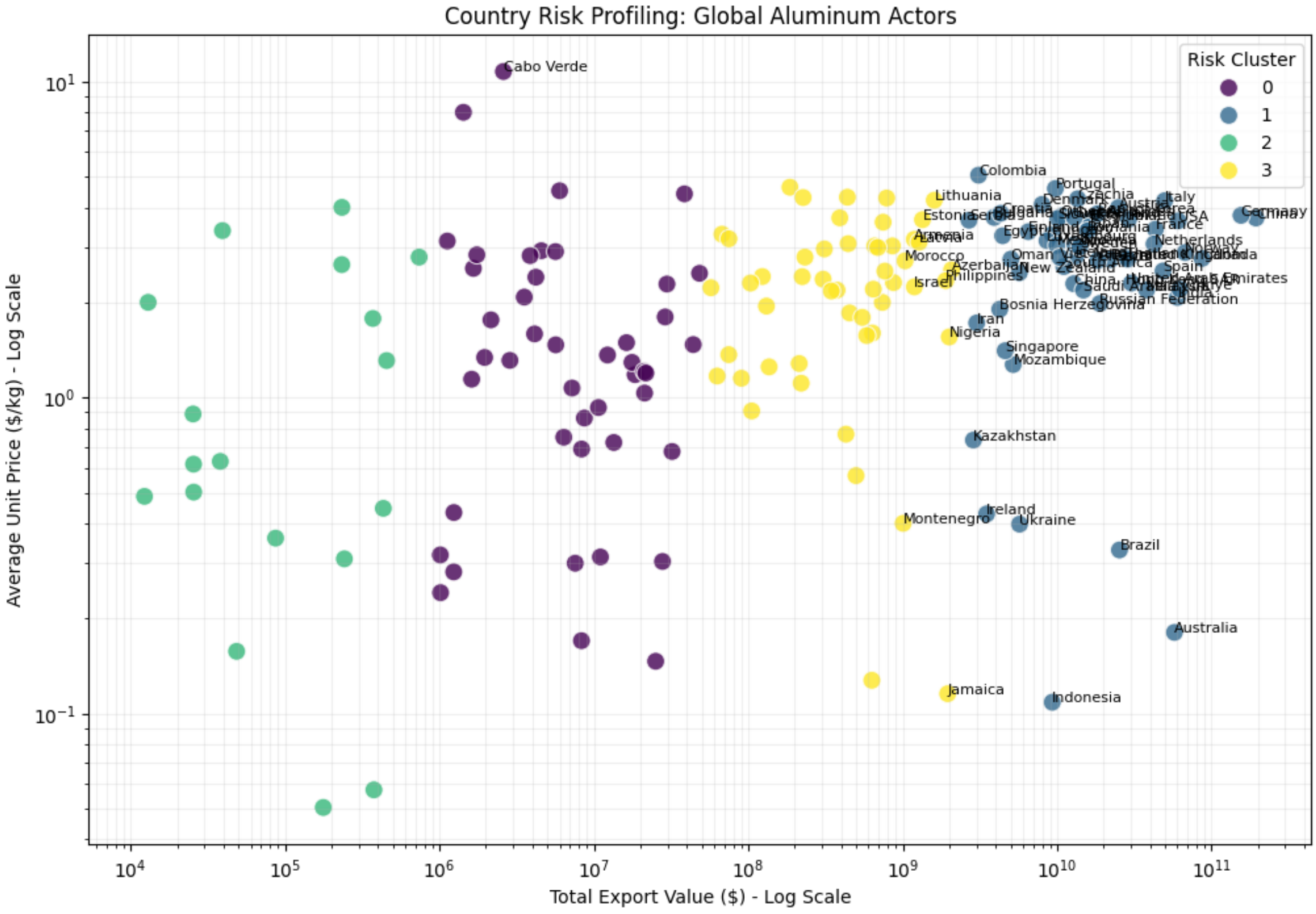

Source: Processed by Author (2025)

**Figure 37. Scatter Plot of Risk Centrality vs. Export Volume Identifying Shadow Hubs.**

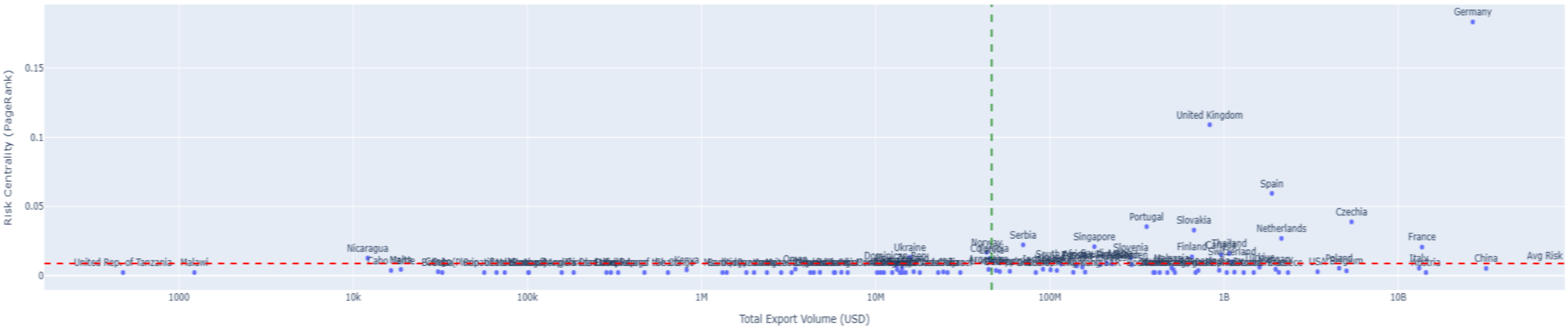

Source: Processed by Author (2025)

This distinction is sharpened by the "Spider Index" in Figure 37, which compares "Risk Centrality" against trade volume. It successfully separates "Major Traders" (bottom right) from "Shadow Hubs" (top left). This reveals that the most dangerous actors are often not the largest exporters, but the pivotal nodes with high network centrality that connect illicit supply chains. By visualizing the "Spider Index," researchers can identify intermediaries that may not move massive volumes but are statistically central to the network of anomalies. To contextualize where these risks sit within the broader market, Figure 38 provides a "Trade Intensity Matrix". This heatmap acts as a map of the "Global Trade Highways," highlighting the dominant corridors where the vast majority of aluminium flows (darker red squares indicate higher volume). By contrasting the "Shadow Hubs" from Figure 37 with the "Legitimate Highways" in Figure 38, we can confirm that fraud is often relegated to the thinner, less visible "backroads" of the global economy, while the massive super-highways such as those connecting major industrial powers in East Asia and Western Europe remain largely consistent in their volume patterns.

**Figure 38. Heatmap of Transaction Volumes Between Top Exporters and Importers.**

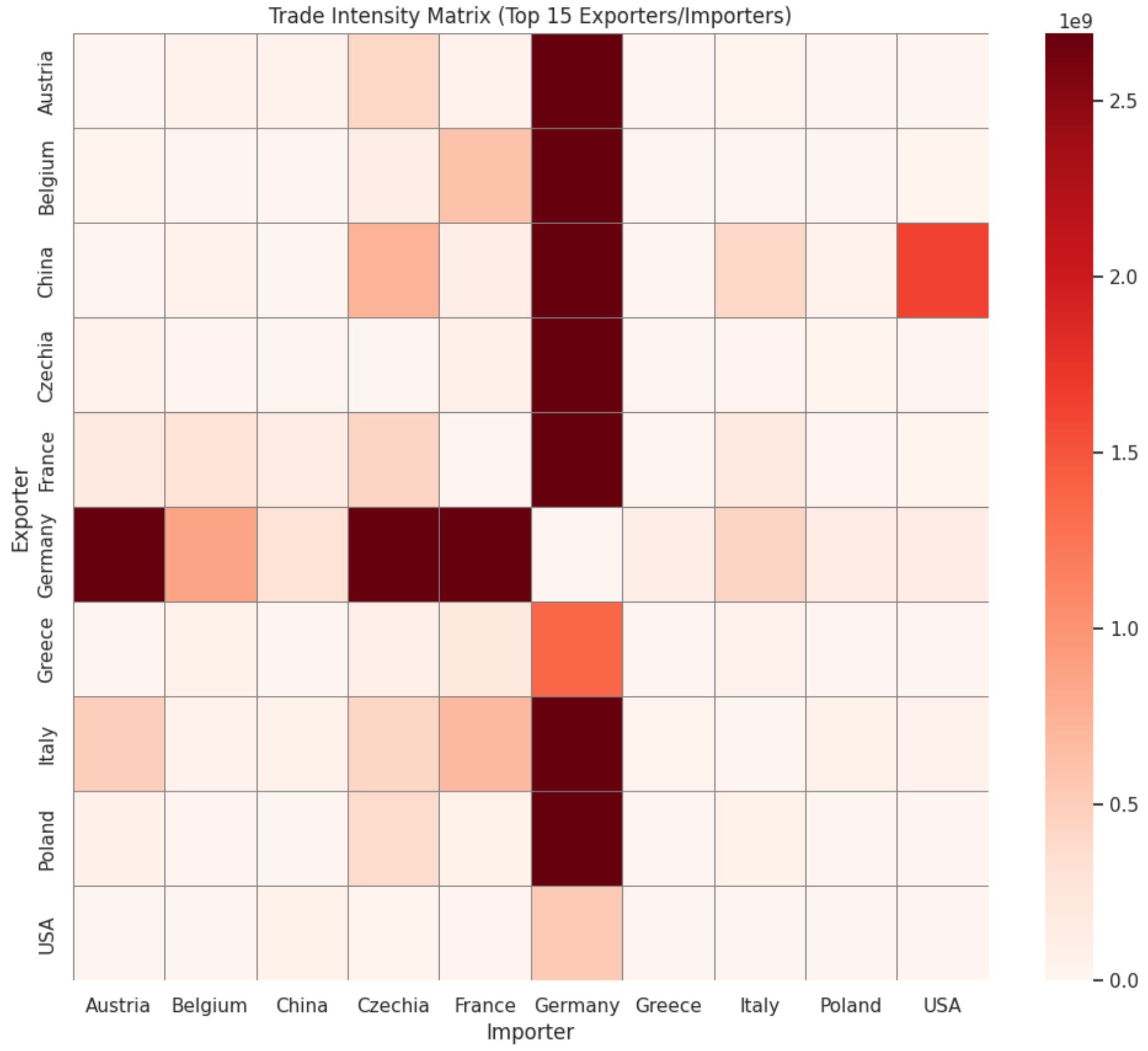

Source: Processed by Author (2025)

### 5.3 The Risk Fingerprint

Finally, we characterize the specific "shape" of a risk event. Figure 39 utilizes a radar chart to generate a multivariate "fingerprint" for anomalies. Unlike normal trades, which are represented by the small, centered blue polygon, anomalies (the red line) exhibit a spiked, irregular shape. This visualization demonstrates that a fraudulent transaction distorts multiple dimensions simultaneously, specifically "Log_Gap_Ratio" and "Price_Disparity", creating a distinct geometric signature that differs fundamentally from the symmetrical profile of legitimate trade.

**Figure 39. Multivariate Risk Radar of Anomaly Fingerprints.**

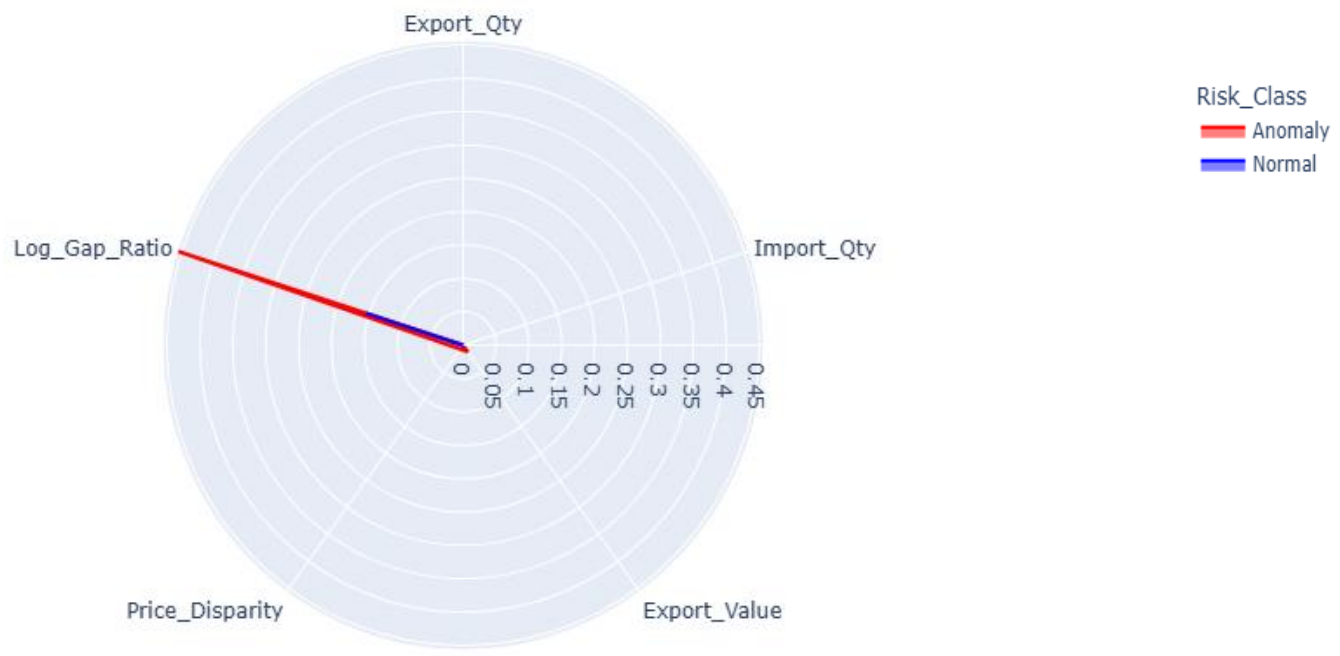

Source: Processed by Author (2025)

**Figure 40. Average Risk Score Analysis by HS Code Commodity.**

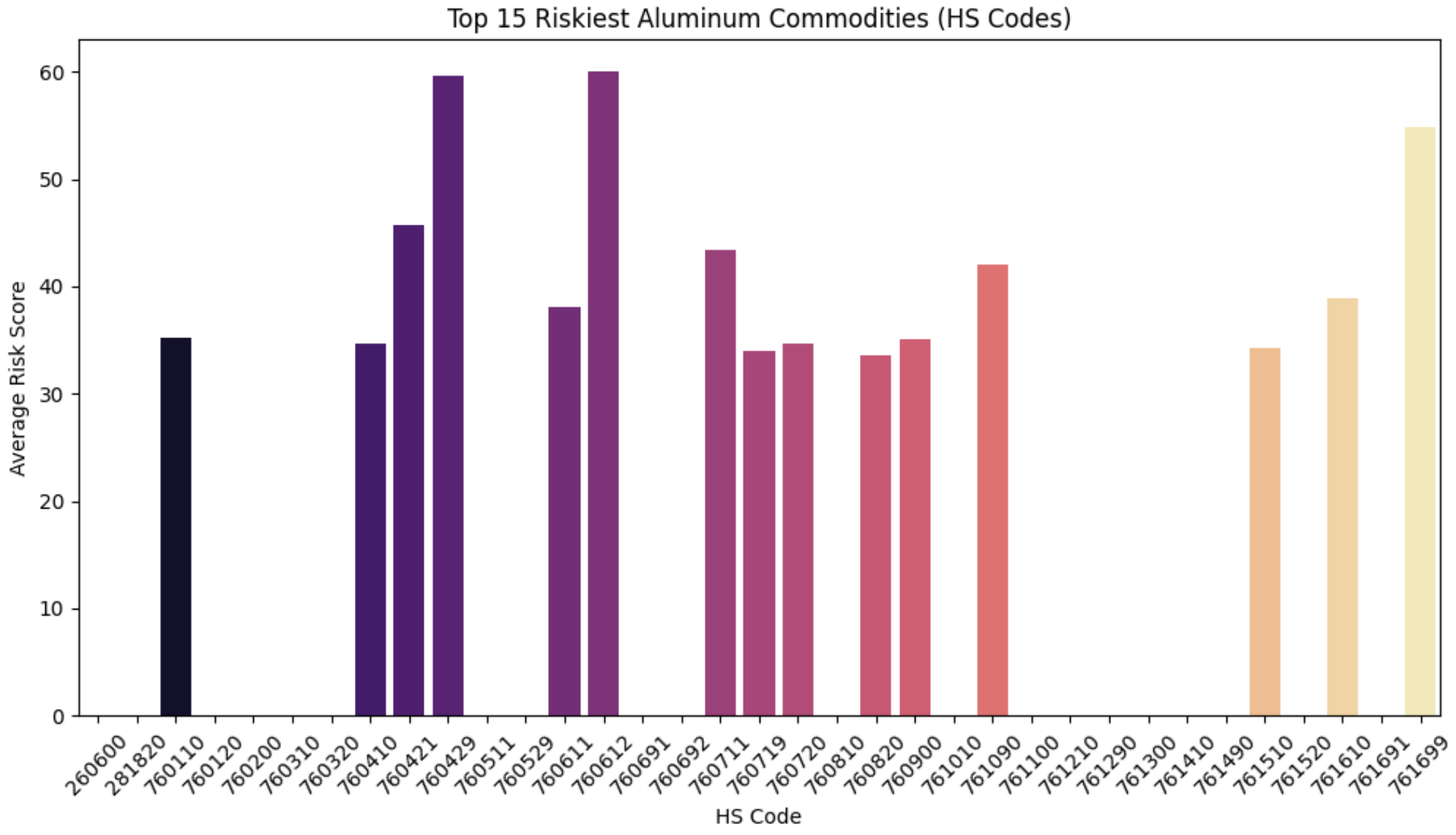

Source: Processed by Author (2025)

This finding is supported by the HS Code Risk Analysis in Figure 40, which identifies specific commodities prone to fraud. The bar chart highlights that high-duty codes (like 7604 and 7606) carry significantly higher risk scores (tall purple bars) compared to raw materials. This validates the "Tax Wedge" theory, confirming that fraud is not distributed evenly across the market but is heavily concentrated in specific product lines where the financial incentive to evade duties is highest. The collective findings, visually confirmed by the clear statistical separation in the "Price Physics Breach", reveal that modern trade fraud has mutated from a physical crime of concealment into a rationalized form of financial arbitrage. The data demonstrates that the primary mechanism is now extreme price manipulation rather than volumetric smuggling, evidenced by high-risk clusters trading at markups which are statistically distinct from normal market behavior. This reality exposes a counter-intuitive network topology where the most significant risks emanate not from high-volume "Major Traders" but from highly central "Shadow Hubs" operating in opaque corridors to exploit specific fiscal "Tax Wedges". Consequently, the adversary is revealed as a rational economic actor whose illicit footprint is defined by a distinct, asymmetrical geometric signature, a "spiked" distortion of value, proving that fraud is no longer a game of hide-and-seek with physical containers, but a sophisticated exploitation of valuation gaps and network connectivity.

## 6. Conclusion

This empirical investigation validates a tripartite taxonomy of trade evasion, confirming that the "tax wedges" created by diverging regulatory regimes are the primary architects of modern trade fraud. The study substantiates that these regulatory pressures are not merely theoretical but act as the causal force driving "Sustainability Arbitrage" (the greenwashing of scrap), catalyzing "Fiscal Asymmetries" (specifically the "Mirror Fracture" in trade data), and necessitating "Geopolitical Rerouting" to bypass sanctions. However, the analysis uncovers a critical operational vulnerability beyond these drivers: the strategic use of "Void-Shoring," a suppression maneuver where illicit actors sever the forensic link between export and import records by routing assets into a statistical "void" rather than a physical jurisdiction. This mechanism effectively blinds regulators to the true trajectory of high-risk assets, creating a "Black Hole" in global trade data.

To dismantle these obfuscation mechanisms, customs authorities must transition from static, rule-based monitoring to a "Glass Pipeline" Protocol, a dynamic dual-layer detection system designed to illuminate the dark corners of the supply chain. The first pillar, "Dynamic Price Banding," serves as the primary defense against TBML. Validated by SHAP analysis which identified "Price_Deviation" as the dominant predictor of fraud, this system replaces fixed thresholds with continuous benchmarking against the global median. Transactions that breach this dynamic standard deviation such as the identified "Hardware Mask" anomalies trading at markups exceeding 1,900% (approx. $167/kg), would trigger an immediate "Audit Hold," allowing authorities to intercept over-invoicing schemes before the capital flight is finalized.

To address the network vulnerabilities, customs agencies must implement a Mandatory Destination Reconciliation protocol. This measure is specifically designed to dismantle "Null-State Logistics", the widespread practice of clearing high-volume shipments via "Unspecified" (Code 0) fields to erase the cargo's digital footprint. Our analysis revealed that major exporters, including the UK and Norway, utilized these "Ghost Routes" to shield high-value anomalies from mirror analysis. By enforcing a strict "Mirror Validation" requirement that mandates valid ISO country codes for clearance, authorities can force trade flows out of the statistical grey zone, neutralizing the "Ghost Destination" mechanism and re-enabling the use of mirror statistics for forensic investigations.

Finally, while this framework was calibrated using aluminium trade data, the arbitrage mechanics identified herein represent a systemic risk to the wider resource economy. As the global energy transition accelerates, the rising "Greenium" and demand for critical minerals will likely exacerbate these market distortions. Future research must therefore extend this unified machine learning framework to the Copper and Lithium sectors, as these supply chains are equally vital to the decarbonization agenda and uniquely susceptible to the opacity risks described in this study.